%% file: plstudy.tex
\documentclass[10pt,journal,compsoc]{IEEEtran}
\pdfoutput=1

\ifCLASSOPTIONcompsoc
  \usepackage[noadjust]{cite}
\else
  \usepackage{cite}
\fi

\ifCLASSINFOpdf
\else
\fi

\usepackage{cite}
\usepackage{CJKutf8}
\usepackage{graphicx}
\usepackage{textcomp}
\usepackage{xcolor}
\usepackage{tcolorbox}
\usepackage{booktabs} %
\usepackage[ruled,linesnumbered]{algorithm2e}
\usepackage{booktabs}
\usepackage[british]{babel}
\usepackage{paralist}
\usepackage{multirow}
\usepackage{amsmath}
\usepackage{amssymb}
\usepackage[marginal]{footmisc}
\usepackage{subfigure}
\usepackage{balance}
\usepackage{threeparttable}
\usepackage{hyperref}
\usepackage{color}
\usepackage{colortbl}
\usepackage{array}
\usepackage[framemethod=default]{mdframed}
\usepackage{showexpl}
\usepackage{comment}
\usepackage{float}
\usepackage{url}
\makeatletter
\g@addto@macro{\UrlBreaks}{\UrlOrds}
\makeatother
\usepackage{framed}
\usepackage{alltt}
\usepackage{enumitem}

\SetCommentSty{mycommfont}
\usepackage{algpseudocode}
\usepackage{tikz}
\usetikzlibrary{trees}
\usepackage{pgfplots}
\usepackage{bchart}
\usepackage{makecell}
\usepackage{caption} 
\newcommand{\ploc}{\emph{patchloc}}

\newcommand{\ptime}{\emph{resolutiontime}}

\newcommand{\python}{Python}
\newcommand{\cpp}{C++}
\newcommand{\csharp}{C\#}
\newcommand{\ruby}{Ruby}
\newcommand{\php}{PHP}
\newcommand{\obc}{Objective-C}
\newcommand{\js}{JavaScript}
\newcommand{\go}{Go}
\newcommand{\cc}{C}
\newcommand{\java}{Java}

\newcommand{\jie}[1]{{\color{black} #1}}
\newcommand{\jienew}[1]{{\color{black} #1}}
\newcommand{\pleasecheck}[1]{{\color{black} #1}}

\def\BibTeX{{\rm B\kern-.05em{\sc i\kern-.025em b}\kern-.08em
    T\kern-.1667em\lower.7ex\hbox{E}\kern-.125emX}}

\begin{document}

\title{A Study of Bug Resolution Characteristics in Popular Programming Languages}

\author{Jie M. Zhang, Feng Li, Dan Hao, Meng Wang, Hao Tang, Lu Zhang, Mark Harman
\IEEEcompsocitemizethanks{\IEEEcompsocthanksitem Jie M. Zhang is the corresponding author. She has two affiliations, Peking University and University College London, of which, for this paper, the primary affiliation is with the Key Laboratory of High Confidence Software Technologies (Peking University), MoE, China. She is currently with UCL, but this work was started when she was a PhD student at Peking University. Feng Li, Dan Hao, Hao Tang, and Lu Zhang are with the Key Laboratory of High Confidence Software Technologies (Peking University), MoE, China.  \protect }
 \IEEEcompsocitemizethanks{\IEEEcompsocthanksitem Mark Harman is with University College London and Facebook London, United Kingdom.\protect}
 \IEEEcompsocitemizethanks{\IEEEcompsocthanksitem Meng Wang is with the Department of Computer Science, University of Bristol, UK.\protect\\}

}

\markboth{}%
{Zhang \MakeLowercase{\textit{et al.}}: Bare Advanced Demo of IEEEtran.cls for IEEE Computer Society Journals}

\IEEEtitleabstractindextext{%
\begin{abstract}

This paper presents a large-scale study that investigates the bug resolution characteristics among popular Github projects written in different programming languages.
We explore correlations but, of course, we cannot infer causation.
Specifically, we analyse bug resolution data from approximately 70 million Source Line of Code, drawn from 3 million commits to 600 GitHub projects, primarily written in 10 programming languages.
We find notable variations in apparent bug resolution time and patch (fix) size. 
While interpretation of results from such large-scale empirical studies is inherently difficult, we believe that the differences in medians are sufficiently large to warrant further investigation, replication, re-analysis and follow up research.
For example, in our corpus, the median apparent bug resolution time (elapsed time from raise to resolve) for Ruby was 4X that for Go and 2.5X for Java.
We also found that patches tend to touch more files for the corpus of strongly typed and for statically typed programs.
However, we also found evidence for a {\em lower} elapsed resolution time for bug resolution committed to projects constructed from statically typed languages.
These findings, if replicated in subsequent follow on studies, may shed further empirical light on the debate about the importance of static typing.
\end{abstract}

}

\maketitle

\IEEEdisplaynontitleabstractindextext

\IEEEpeerreviewmaketitle

\input{Sections/all_in_one.tex}

\bibliographystyle{unsrt}
\bibliography{sample-bibliography}

\begin{IEEEbiography}[{\includegraphics[width=1in,height=1.25in,clip,keepaspectratio]{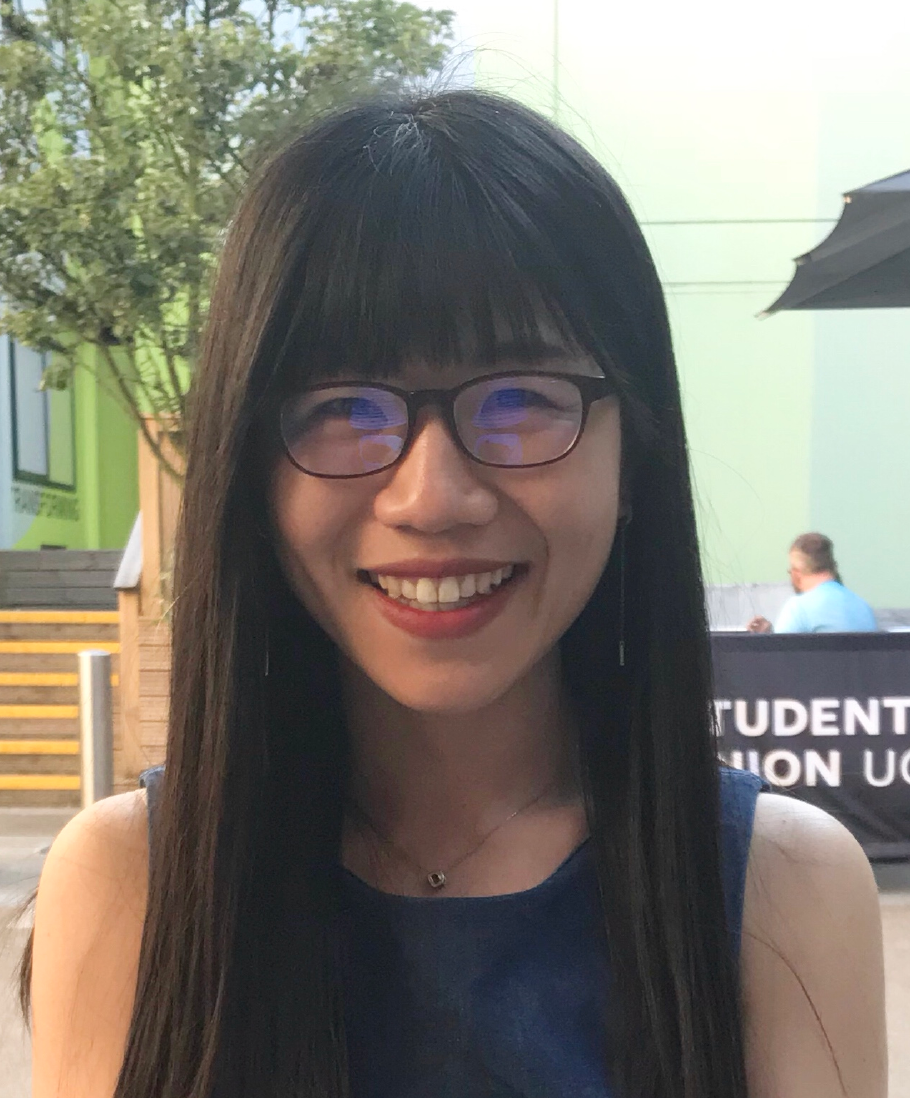}}]{Jie M. Zhang}
is a research fellow at CREST, UCL. She works with Prof. Mark Harman. She got her Ph.D degree at Peking University in 2018. She has won the 2016 MSRA Fellowship, the Top-ten Research Excellence Award of EECS, Peking University, and the 2015 National Scholarship. She is the co-chair of ASE SRC 2019 and Mutation 2020, and the program committee member of FSE 2020, ISSTA 2020, ASE 2019. Her major research interests are software testing and machine learning testing. 
\end{IEEEbiography}	
\begin{IEEEbiography}[{\includegraphics[width=1in,height=1.25in,clip,keepaspectratio]{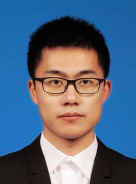}}]{Feng Li}
received his B.S. degree from Peking University in 2018. He is currently a PhD student in the School of Electronics Engineering and Computer Science of Peking University. His research interests include software testing and analysis.
\end{IEEEbiography}	

\begin{IEEEbiography}[{\includegraphics[width=1in,height=1.25in,clip,keepaspectratio]{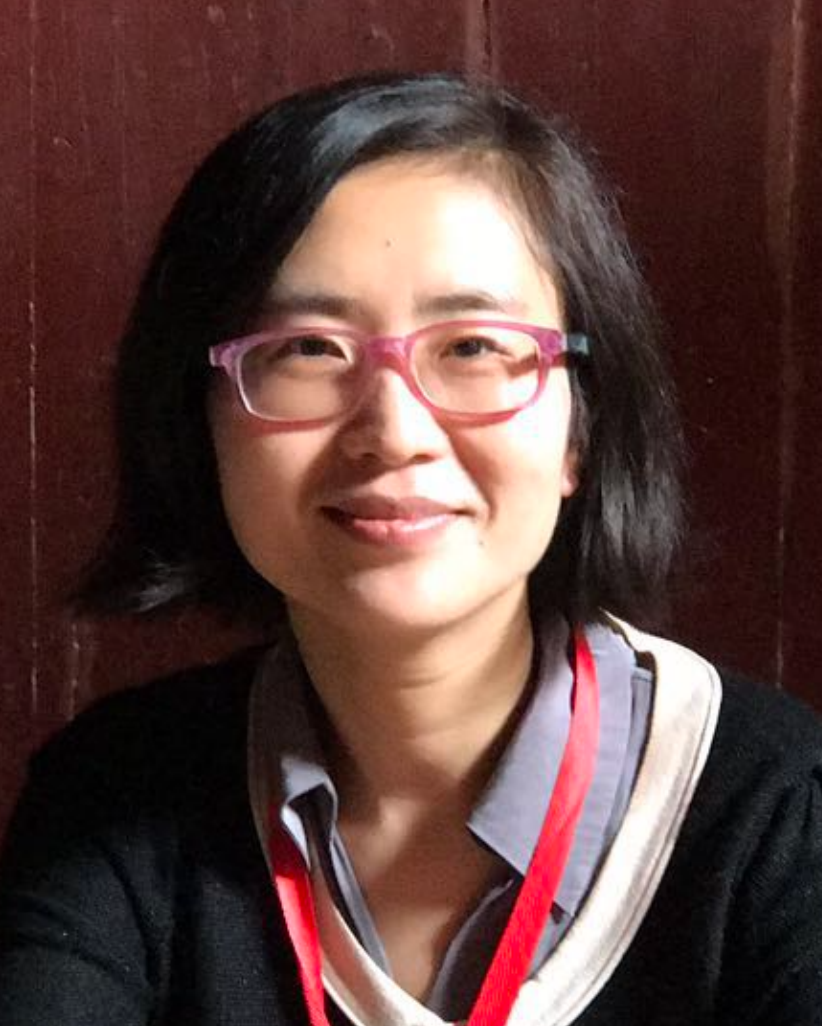}}]{Dan Hao}
is an associate professor at EECS, Peking University, P.R.China. She received her Ph.D. in Computer Science from Peking University in 2008, and the B.S. in Computer Science from the Harbin Institute of Technology in 2002.  She was a general co-chair of SPLC 2018, in the organisation team of ICST 2017, ICST 2019 and ISSTA 2019, the program committees of many prestigious conferences such as ICSE and ASE. Her current research interests include software testing and debugging.
\end{IEEEbiography}	
\begin{IEEEbiography}[{\includegraphics[width=1in,height=1.25in,clip,keepaspectratio]{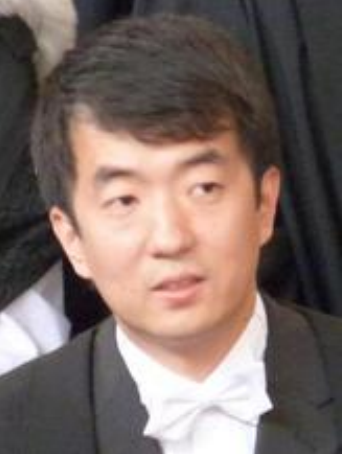}}]{Meng Wang} is a senior lecturer at University of Bristol after faculty level appointments at University of Kent and Chalmers University of Technology. The central theme of his research is to apply theoretical rigour to practical programming, with the aim of improving the correctness and robustness of software systems. In particular, he focuses on designing languages and tools for software development and testing. He publishes regularly at top conferences in programming languages including ICFP and OOPSLA, and journals including Journal of Functional Programming and Science of Computer Programming. He serves as general co-chair for the Symposium on Trends in Functional Programming and is a steering committee member. He contributed to the program committees of more than 15 ACM and LNCS conferences and workshops, and he acts as referee for major journals. He is the PI of four projects funded by the Royal Society and EPSRC. 
\end{IEEEbiography}
\vspace{-1.3cm}
\begin{IEEEbiography}[{\includegraphics[width=25mm,height=32mm,clip,keepaspectratio]{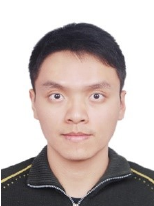}}]{Hao Tang}
received his Ph.D degree in Computer Science from Peking University in 2019. He has won the 2015 MSRA Fellowship. His research interests include static analysis and defect prediction.
\end{IEEEbiography}	

\begin{IEEEbiography}[{\includegraphics[width=1in,height=1.25in,clip,keepaspectratio]{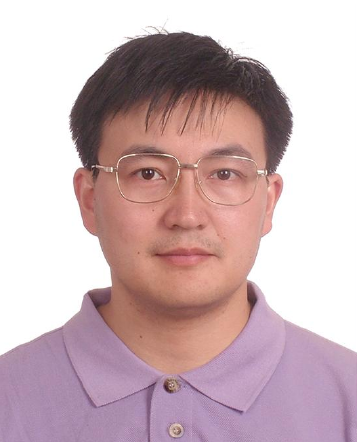}}]{Lu Zhang}
is a professor at EECS, Peking University, P.R. China. He received both PhD and BSc in Computer Science from Peking University in 2000 and 1995 respectively. He was a postdoctoral researcher in Oxford Brookes University and University of Liverpool, UK. He served on the program committees of many prestigious conferences, such FSE, OOPSLA, ISSTA, and ASE.  He was a program co-chair of SCAM2008 and a program co-chair of ICSME17. He has been on the editorial boards of Journal of Software Maintenance and Evolution: Research and Practice and Software Testing, Verification and Reliability. His current research interests include software testing and analysis, program comprehension, software maintenance and evolution, software reuse and component-based software development, and service computing
\end{IEEEbiography}	
\begin{IEEEbiography}[{\includegraphics[width=25mm,height=32mm,clip,keepaspectratio]{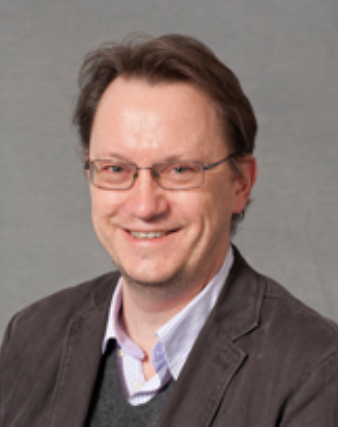}}]{Mark Harman}
works full time at Facebook London as a Research Scientist in the London Probable $\langle$T$\rangle$ Team. The team seeks to incubate and deploy research that has potential to transform approaches to Reliability, Integrity and Privacy. He also holds a part-time professorship at UCL. Previously, Mark was the manager of the Facebook team that deployed Sapienz to test mobile apps, leading to thousands of bugs being automatically found and in multimillion line communications and social media apps in daily use by over 1.4Bn people worldwide. Sapienz grew out of the research spin-out Majicke, which was co-founded by Mark and two of his UCL colleagues and which was acquired by Facebook in 2017. In his more purely scientific work, Mark co-founded the field Search Based Software Engineering (SBSE), a research area with over 1,000 authors spread over more than 40 countries worldwide. He is also known for scientific research on source code analysis, software testing, app store analysis and empirical software engineering. He received the IEEE Harlan Mills Award and the ACM Outstanding Research Award in 2019 for this work. In addition to Facebook itself, Mark’s scientific work is also supported by the European Research Council (ERC), with an advanced fellowship grant, and has also been supported by the UK Engineering and Physical Sciences Research Council (EPSRC), for example, with larger, longer-term platform and programme grants.
\end{IEEEbiography}	
\end{document}

%% file: Sections/all_in_one.tex
\section{Introduction}
\label{sec:introduction}
For several decades there have been great debates about the influence of programming language paradigm choices (such as declarative/imperative or strong/weak typing) on software engineering concerns (such as bug prevalence and resolution characteristics).
For the most part, such debates have been largely uninformed by empirical scientific evidence, having more the character of anecdotal opinion sharing rather than thorough scientific evidence-based decision-making.
However, in 2014, Ray et al.~\cite{ray2014large} \jie{began a project of reformulating such questions as} large-scale empirical analyses.
In particular, they \jie{addressed} one of the most immediate and pressing questions: `which programming language paradigms are more correlated with software quality?'.

There is a fundamental empirical spectrum, along which researchers must make methodological choices when answering such `great debate' questions.
At one end of the spectrum lies the approach used by Ray et al.~\cite{ray2014large}, which sacrifices some degree of control over variables for greater scalability and increased probability of generalisability.
At the other end of the spectrum lie specific, focused and more `controlled' studies that seek to, at least partly, account for tools, environments, ecosystems and engineers involved in constructing software, thereby removing some potential confounding factors.
Such studies trade scale and generalisability for greater control of variables that might otherwise influence the results observed.
There are several previous studies~\cite{bhattacharya2011assessing,kleinschmager2012static,hanenberg2014empirical} that lie at this end of the spectrum methodological choices.

Since the essence of any `great debate' question about programming languages and software engineering tends to rest on generalisations, any attempt at an answer is inherently pushed towards the large-scale/less controlled end this spectrum of methodological choices. %
As a result, one can only speak of correlations observed \jie{between corpora developed within} the overall ecosystems that surround the sets of programs and systems studied in such corpora.

Any such large scale study of systems in many languages draws on data extracted from code repositories for which, due to scale, analysis will be necessarily partly automated, labelling will be partly subjective and data will be, itself, partial and noisy. Consequently, there will be many inherent potential treats to the validity of any scientific conclusions. 
Actionability, for any {\em initial} study such as this, will also be limited to suggesting the need for follow-on replication and further and deeper study of some of the key observations. 
If such empirical observations  prove to hold up under replication and re-analysis, then progress then can be made towards shedding light on great debates.
\begin{CJK*}{UTF8}{gbsn}
As the Chinese proverb has it, 
``千里之行，始于足下''; 
\end{CJK*}
a journey of a thousand miles begins with a first step.

In this paper we focus on the question: ``{\it for projects constructed from popular programming languages in widespread current use, which sets of projects exhibit more correlations with longer bug resolution time and larger size of fix (or patch)?''. We also explore correlations between bug resolution characteristics and other aspects, such as categories of approaches to typing (strong, weak, static, dynamic), project features (such as overall age and size and domain of application). }

Opinions and anecdotes about the importance (or otherwise) of types and type systems have raged and ranged over several decades.
For example, some have argued that static typing tends to result in better software quality and lower bug-resolution cost because type checking is \jienew{deemed to be} an effective way of catching bugs earlier~\cite{dynbadone,dynbadtwo}, while others
have argued that dynamic typing eases reading, writing, and understanding, thereby, so the argument goes, making code constructed from such languages easier to debug~\cite{nierstrasz2005revival}.
\jienew{Many} more thorough empirical studies are required to move the debate onto a firmer scientific footing.

Our starting point was the large-scale/less controlled end the spectrum reminiscent Ray et al.~\cite{ray2014large}.
We refined and augmented their analysis approach (Section~\ref{sec:methodology}).
Given the inherent difficulties of large scale analysis there are many threats to validity, and our methodology is, consequently, surely far from perfect. 

In the remainder of the paper, we have attempted to highlight potential  threats to the validity of our observations and conclusions, throughout the paper as well as in the traditional `Threats to validity` section (Section~\ref{threats}).
Where practical, we have taken steps to explore potential sources of bias, mislabelling, miscatagorisation, noise and confounding factors.
Nevertheless, it is likely that we have overlooked others, and so we make our data publicly available to support subsequent authors in re-analysis and re-evaluation of both our observations and the conclusions drawn from them.

We analysed bug-resolution data from approximately 70 million Source Lines of Code (SLOC) drawn from 3 million commits to 600 GitHub projects in 10 languages.
We adopted a variety of measurement criteria to investigate the data and provided empirical evidence to support the scientific conclusions drawn from the data.
Specifically,
\begin{enumerate}
    \item We report results from multiple statistical analyses (median-value comparison, multiple regression, and ScottKnott analysis. 
    \item Some of our primary observations of note are made using median values (of a large number of commits and issue reports), which is comparatively simple and intuitive.
    Median values also tend to be `less affected by outliers and skewed data'~\cite{bissyande2013got,wonnacott1972introductory,meanandmedian};
    \item We manually inspected the data we analysed to \jienew{investigate and account for possible mislabelling problems arising from} automatic data extraction.
\end{enumerate}

Our findings reveal notable differences in bug-resolution size and time between individual languages and language categories.
In particular, we found evidence that projects in our corpus constructed from  \java{} and \csharp{} tended to involve larger fix sizes than the other languages during bug resolution.
This finding, of replicated in subsequent studies, may inform automated repair~\cite{arcuri2008novel,hansson2015automatic,arcuri2008automation}, which may need to search larger spaces for candidate bug fixes.
It may also inform mutation testing~\cite{jia2010analysis,papadakis2019mutation,zhang2019pseudo,zhang2014empirical}, which may need more code transformations to simulate real faults. 
Furthermore, projects in our corpus constructed from \java{}, \go{}, and \python{} tended to consume less bug-resolution time.
Systems constructed from statically/strongly typed languages tended to involve larger modification sizes to resolve bugs.
We also found that systems in our corpus constructed from weakly/dynamically typed languages tended to have longer bug-resolution time.
The main contributions of this paper are as follows:

\noindent \textbf{1) A large scale empirical study} on bug-resolution characteristics among Github projects written in different programming languages and categories. We perform a  study of 10 popular programming languages.

\noindent \textbf{2) Empirical evidence} of notable differences in median bug resolution time and size between the programs in our corpus. Most notably:
\begin{enumerate}[label=\alph*)]
\item \pleasecheck{\java{} and \csharp{} bug-resolutions change 2.2 times as many lines as \ruby{}. }

\item \java{} and \csharp{} bug-resolutions touch 2 times as many files as \ruby{} and \python{}. 

\item \pleasecheck{\obc{} bug-resolution consumes 4.6/3.0 times as much bug-resolution time as \go{} in project/commit-level analysis; \ruby{} consumes 2.5/8.1 times as much bug-resolution time as \java{} in project/commit-level analysis;}
\end{enumerate}

\noindent \textbf{3) Potentially interesting differences} between categories of programs within our corpus, based on their approach to typing. For example, in our project-level analysis: 
\begin{enumerate}[label=\alph*)]
\item Dynamically typed language bug-resolutions change 37.5\% fewer lines, yet consume 59.5\% more bug-resolution time than statically typed language bug resolutions. 

\item Weakly typed language bug-resolutions change 20\% fewer lines, yet consume 46.5\% more bug-resolution time than strongly typed language bug resolutions.
\end{enumerate}

In Section~\ref{implication}, we discuss some possible implications for other software engineering research, from  these findings.

\section{Methodology}
\label{sec:methodology}
In this section, we describe how we \jienew{extract} bug-resolution characteristics ( Section~\ref{sec:criteria}) and what statistical methods we adopted to conduct the study (Section~\ref{sec:analysisapproach}).

\subsection{Bug-Resolution Size and Time} 
\label{sec:criteria}
We consider bug-resolution size from two aspects to better understand the dispersion degree of a bug fix: \pleasecheck{1)~Sloc changed: the number of modified executable lines, which indicates the amount of code that requires modification to fix a bug; 2)~Files touched: the number of modified files.}  
In addition to size, we also study bug-resolution time, i.e., the time lapse between the reporting and resolution of bug. 
Zheng et al. \cite{Zheng:2015:MIC:2786805.2786866} provided empirical evidence that the timestamp of the last commit \pleasecheck{before issue closing} is the most reliable assessment of resolution time from the available alternatives, so we adopt this during our measurement.

For each project, we collect bug-resolution size by analysing buggy commits (more details in Section~\ref{ExperimentalSetup}). 
For an experiment of this scale, such collection has to be automatic. 
We favour precision to recall when selecting bug-resolution commits, considering that non-bug-resolution commits may bring \jienew{additional yet avoidable} bias to the data. 
In particular, we identified `fix' and `bug' as being the best keywords for searching,
which are least exposed to false positives. This choice is confirmed by our manual analysis (achieving 95\% precision), whereas other candidate keywords `issue', `mistake', or `fault' achieved only approximately  30\% precision\footnote{These keywords may also be widely used to describe problems unrelated to source code bugs (e.g., issues in documents), which has been cited as one potential source of potential mislabelling~\cite{attackpaper}.}. 
Once a bug-resolution commit is identified, we extract the number of modified program files and lines belonging to the project's primary language (more details in Section~\ref{sec:procedure}).

When a bug is fixed, a range of files may be modified/updated. 
In our measurement, we exclude non-code modifications such as documentation and comments but count all code changes of both source and test programs. 
This choice is deliberate, as we believe developer testing is an integrated part of development~\cite{zhang2016isomorphic,zhang2018predictive}, and the effort involved in updating test code is naturally part of bug resolution and is language dependent. 

One threat to the validity of this approach is that some bug-resolution commits may also contain code modification unrelated to the bug, such as refactorings which are likely to affect a disproportionally large amount of code~\cite{zhang2018automated}. 
To check the severity of this \jienew{potential source of} bias, we manually analysed 520 randomly-chosen bug-resolution commits from all our projects and found that 14.6\% of commits involved dealing with more than a single bug or other forms of code modifications unrelated to bug resolution, indicating a \jienew{reasonable}, yet not unassailable,  level data integrity\footnote{\jie{The manual analysis results are on our project homepage~\cite{thispaperhomepage}}.}.

To further reduce potential bias, for each project, we used the median value of all the bug-resolution commits to represent the project's general level of line/file modification in project-level analysis, which tends to be `less affected by outliers and skewed data'~\cite{bissyande2013got}.

We acquire the bug-resolution time for each project by analysing issue reports~\cite{bissyande2013got,Zhang2015}. We do not use commit information here, because it gives us only the end time but not the corresponding start time of bug-resolution\footnote{When extracting bug-resolution size, we do not use issue reports because we found that only around 50\% of the issue reports are linked with commits.}.
Instead, we search the issue tracking system for closed issues with labels containing
`bug' (case insensitive), and extract
information from them.
In Github, project issues are manually labelled by the developers. 
We treated this as a ground truth (another possible avenue for subsequent re-analysis to assess any potential threats to validity arising from such an assumption).

Inspired by the work of Zheng et al.~\cite{Zheng:2015:MIC:2786805.2786866}, we define the resolution time of each bug as the interval between issue creation and the last comment before issue closing, which has been demonstrated to be more accurate (than the interval between creation and closing time)~\cite{Zheng:2015:MIC:2786805.2786866}. 
Again, we use the median of the so-computed bug-resolution times as a representation of the \jienew{overall project level} bug-resolution time to remove potential bias due to extreme values.

\subsection{Statistical Analysis Used in the Study}
\label{sec:analysisapproach}
We collect the values of dependent variables (bug-resolution size/time),
independent variables (languages),
and four project features (project size, age, number of contributors\footnote{\pleasecheck{Contributors are not necessarily developers~\cite{attackpaper}.}} and commits)
for each project.

We focus on comparison of median values.
However, following Ray et al.~\cite{ray2014large}, we also report results for multiple regression analysis.
Furthermore, following the request of an anonymous referee, we complement this with a report of ScottKnott analysis \pleasecheck{(to indicate difference significance).
These three approaches may mutually validate each other
from different aspects when all produce similar outcomes.}

\pleasecheck{
For each analysis approach, we use two types of data. 
The first type treats each project as an individual. 
During this analysis, each data point concerns a single project.
Each project has its own median bug-resolution size/time among all its commits, which represents this project's overall bug-resolution size/time. 
The second type treats each commit for a given language as an individual. 
During analysis, each data point denotes a single commit.
For ease of presentation, we represent the analysis for first type as \textbf{project-level analysis} and the second type as \textbf{commit-level analysis}.}

\subsubsection{Comparison of Median Values}

We calculate project-level and commit-level median values to represent a language's central tendency of bug-resolution size and time~\cite{srinivasan2007new,bissyande2013got,wonnacott1972introductory,meanandmedian}. 
For project-level analysis, we use the median bug-resolution size/time of all projects belonging to a language to represent this language's central bug-resolution size/time.
For commit-level analysis, we use the median bug-resolution size/time of the whole set of commits from all projects in a language to present this language's central bug-resolution size/time.

To help visualise, we present box plots, which contain the $25^{th}$, $50^{th}$, and $75^{th}$ percentiles in the distribution of values.
We also manually analysed scatter plots for each language and each treatment (size in lines of code and files, and duration). Space does not permit us to include all 30 resulting scatter plots in the paper, but we make these available on the companion website~\cite{thispaperhomepage} for others to investigate.

\subsubsection{Multiple Linear Regression}
We use multiple linear regression to indicate the contribution of different (categories of) languages to the bug-resolution characteristics\footnote{We use the most common `dummy coding'~\cite{bech2005effects} for categorical language variables.}. 
\jie{The comparison among bug-resolution characteristics of different languages (and categories) can be regarded as an importance-determination problem of categorical variables, and thus 
multiple linear regression can be used to identify which (category of) languages contribute more to the bug-resolution size/time~\cite{freedman2009statistical,vittinghoff2011regression}.

Through multiple linear regression of the language variable and bug-resolution size/time of each project or commit, each categorical variable has a regression coefficient, which represents the mean change in the response given a one deviation change in the regression model. 
Higher coefficients indicate larger means of bug-resolution size/time. }
In addition to coefficients, we also present the following statistics: 1)~p-value;
2)~t-value: a statistical test value that measures the ratio between the coefficient and its standard error~\cite{winer1971statistical}; 
3)~F-statistics: a statistical test value that checks the null hypothesis that all of the regression coefficients are equal to zero.

For p values, we make no correction for multiple statistical testing, and prefer to place little emphasis on claims to significance based on (somewhat arbitrary) choices of statistical significance level.  
Instead, we have chosen to report the raw p values\footnote{When conducting ScottKnott analysis, it requires a specific significance level to perform clustering, for which we use the typical level of 0.05.}, in order to allow the reader to place whatever interpretation he or she deems appropriate on these statistics. 
From the the raw values, it is easy to compute, for example, a conservative Bonferroni correction and/or to assess significance at some chosen level of statistical significance \cite{arcuri:practical}.

Statistical significance (at any chosen level) is increasingly likely to be observed for some comparison, as the number of data points studied increases.
Therefore, we prefer to draw conclusions based primarily of \pleasecheck{median-value differences } observed, using significance test outcomes, less formally (and less ritualistically). 
With this in mind, we use the p values merely as one way to give an indication of the confidence with which we observe that an effect is, indeed,  present, given the size of data we have been able to collect.

\subsubsection{ScottKnott Clustering Analysis}

We then report ScottKnott clustering analysis (abbreviated as `ScottKnott analysis' in this paper). 
The ScottKnott algorithm clusters treatment means into distinct homogeneous groups without any overlapping, thus can indicate which groups are significantly different.

\pleasecheck{
When conducting multiple linear regression and ScottKnott analysis, we perform log transformation of bug-resolution size/time (\pleasecheck{a typical data transformation approach in statistical analysis, which is also adopted in previous work~\cite{ray2014large}}). The log transformation  facilitates a better non-linear fit and a normal data distribution.
}

\section{Experimental Setup}
\label{ExperimentalSetup}

This study is designed to answer four research questions.

\noindent\textbf{RQ1: What are the differences of bug-resolution size/time among \jienew{projects written in} different programming languages?} \\
\noindent\textbf{RQ2: What are the differences of bug-resolution size/time among \jienew{projects written in} different programming language categories?}

The first two research questions aim to 
investigate the differences in bug resolution characteristics among different languages and categories. 
For each research question, we ask three sub-RQs, with each sub-RQ covering the result of one type of analysis approach:  \emph{what is the result when using median-value analysis/multiple-regression analysis/ScottKnott analysis}?

\pleasecheck{This paper does not aim to conduct any form of causal analysis. 
However, to complement our analysis of language and language categories, 
we design the following two research questions to investigate the correlations between bug resolution characteristics and other project features and application domains.
}

\noindent \textbf{RQ3: What evidence is there for correlations between bug resolution characteristics and project features: SLOC, number of commits, age and/or contributor?}\\

\noindent \textbf{RQ4: What evidence is there for correlations between bug resolution characteristics and project domain?}\\

\pleasecheck{These research questions aim to investigate the relative degree of correlation observed in RQ1 and RQ2 with that observed for other characteristics.
}

\subsection{Target Programming Languages}
\label{sec:categories}
\jienew{To select a set of languages from which to choose programs to study,}
we consult several rankings of the most popular languages~\cite{mostpopone,mostpoptwo,mostpopthree,mostpopfour,mostpopfive}, and choose the following 10 (in alphabetical order) as our targets: \cc{}, \csharp{}, \cpp{}, \go{}, \java{}, \js{}, \obc{}, \php{}, \python{}, and \ruby{}, to focus on the more popular languages according to these rankings.

We categorise the programs' languages according to two classification systems following previous work~\cite{ray2014large}, as shown in Table~\ref{tab:langcate}. The \emph{compilation} classification classifies a target language into dynamic or static categories based on whether types are checked dynamically during runtime or statically during compilation. 
The \emph{type} classification classifies a target language into strongly typed and weakly typed based on
whether automatic type conversions are allowed. We sometimes use \emph{static} languages and \emph{dynamic} languages to refer to statically and dynamically typed languages, and use \emph{strong} languages and \emph{weak} languages to refer to 
strongly and weakly typed languages. 
We note that such classifications are somewhat open to interpretation \cite{attackpaper}. 
As such, our choices of category also make a good topic for subsequent re-analysis of our results.

We pay special attention to \python{} during classification. \python{} has two major versions: \python{} 2 and \python{} 3. \python{} 2 is  dynamic, whereas static typing is introduced in \python{} 3. We analyse the source code of all the \python{} projects and find that only four projects contain static types in a small number of files (less than 20\% of files). Therefore, we regard \python{} programs as  dynamically typed, while  paying special attention to the four projects containing static typing, when answering RQ2.  

\begin{table}[t]\small
\centering
\caption{Target programming languages and their categories}
\label{tab:langcate}
\resizebox{.49\textwidth}{!}{
\begin{tabular}{p{1.9cm}|c|c|c|c|c|c|c|c|c|c}
\toprule
\textbf{Category}&\textbf{C}&\textbf{C\#}&\textbf{C++}&\textbf{Go}&\textbf{Java}&\textbf{JS}&\textbf{Objective-C}&\textbf{PHP}&\textbf{Python}&\textbf{Ruby}\\
\midrule
static&$\star$&$\star$&$\star$&$\star$&$\star$&&&&&\\
\cline{2-11}
dynamic&&&&&&$\star$&$\star$&$\star$&$\star$&$\star$\\
\midrule
strong&&$\star$&&$\star$&$\star$&&&&$\star$&$\star$\\
\cline{2-11}
weak&$\star$&&$\star$&&&$\star$&$\star$&$\star$&&\\
\bottomrule
\end{tabular}}
\end{table}

\subsection{Subjects}
\label{sec:subjects}

\begin{table}[t]\small
\centering
\caption{Basic information of projects.}
\label{tab:proinfo}
\resizebox{.49\textwidth}{!}{
\begin{tabular}{lrrrr}
\toprule
\textbf{Language}&\textbf{SLOC}&\textbf{Age (year)}&\textbf{Commits}&\textbf{Contributors}\\
\midrule
\cc{}&2,836--1,036,341&0.45--8.29&7--102,383&1--427\\
\csharp{}&710--5,556,924&0.27--8.53&11--25,543&1--387\\
\cpp{}& 1,115--4,182,449&0.40--8.42&52--44,071&2--435\\
\go{}&920--1,958,917&0.26--7.52&52--47,122&1--468\\
\java{}&323--644,003&0.22--7.34&17--38,379&1--407\\
\emph{JS}& 222--1,311,821&0.81--8.20&130--18,664&10--464\\
\php{}& 161--1,314,420&1.43--7.68&10--52,493&2--432\\
\python{}& 130--723,708&0.19--9.00&89--99,066&3--447\\
\obc{} & 296--179,744&0.50--8.93&70--9,952&4--274\\
\ruby{}& 353--99,250&1.21--9.32&212--63,882&12--443\\
\bottomrule
\end{tabular}}
\end{table}
All our subjects are open-source projects from GitHub~\cite{githubhomepage}. For each target language, we retrieve the project repositories that are primarily written in that language, and select the 60 most popular projects based on their number of stars~\cite{starpopular} as in prior work
~\cite{ray2014large,bhattacharya2011assessing}.
Choosing to focus our study on popular projects (and basing this on star ratings) is, of course, 
another source of potential bias in our sample, that will likely affect the degree to which we can generalise findings beyond our corpus.
There are difficulties in selection bias, no matter how one chooses projects, so generalisations can only be based on multiple such studies.
We encourage others to investigate replication using other choices of sample.

Table~\ref{tab:proinfo} presents the basic information of all the projects. The ranges of four types of information are presented: 
1)~\emph{SLOC}: the physical executable lines of code of the newest version, which is calculated by the tool \emph{CLOC}~\cite{clochomepage}. 
For multi-language projects, we only use the SLOC of the primary language reported by \emph{CLOC}. 
2)~\emph{\#Commit}: the total number of commits downloaded from the GitHub API~\cite{githubapihomepage}. 
3)~\emph{Age}: the age of each project. We use the time interval from creation time recorded to download time as the age (years). 
4)~\emph{\#Contributor}: the number of contributors, which is also collected through GitHub API.

The table shows that the projects of different languages tend to have different project features. 
We study whether these variables correlate with bug-resolution size/time (in RQ3).

\subsection{Experimental Procedure}
\label{sec:procedure}
The experimental procedure of this study can be divided into data collection and data analysis.

\subsubsection{Data Collection}
\label{sec:datacollection}

First, we collect the projects in various programming languages for further analysis.

\emph{Step 1. Information retrieval from GitHub API.} GitHub API provides comprehensive information on commits, issues, and project history\footnote{\jie{We did not use GitTorrent~\cite{GitTorrent} \jienew{to obtain the project information} because that it has not been maintained since 2015.}}. For commits, we download all the \emph{JSON} files of commits, which contain commit messages, the number of line additions and deletions, file changes, and so on. To compute bug-resolution time,
we download the \emph{JSON} files of issues, which contain issue title,
labels, state, creation time, close time, and the times of every comment. \jienew{Due to a restriction of the GitHub API (5,000 accesses} per hour), we skip projects with very large commit
history (which cannot be downloaded within 24 hours)\footnote{In total 15 projects (2.5\%) were skipped due to this constraint.}.

\emph{Step 2. Extraction of related information}. As described in Section~\ref{sec:criteria}, we identify bug-resolution commits through keyword searches. 
Some projects contain multiple languages, for which we only extract changed code belonging to their primary language (\pleasecheck{the language that occupies the most executable lines of code}). 
Specifically, we use Github's own file extension library, the Linguist library\footnote{\url{https://github.com/github/linguist}}, to identify relevant changed files. 

As far as we know, the file extensions in the Linguist library for the 10 languages we studied do not have coincidental mislabelling that may lead to inaccurate language file identification, which has been noted as a possible threat to validity \cite{attackpaper}.

\emph{Step 3. Filtering}. We observe that the `popular' project criterion (based on stars) is also associated  with other indicative metrics such as \#issues, \#contributors, and \#commits. For example, in our subjects, only 1 project has fewer than 10 issues; only 6 have fewer than 20 commits. \pleasecheck{We also manually checked the corpus for duplicate 
projects and found no evidence of duplication. }

We first removed projects with no bug-resolution commit (this resulted in 65 projects being excluded). 
Having applied this filter, the language with fewest remaining projects had 46 projects. 
Therefore, in order that we maintain the same number of projects per language, we chose 46 projects per language from the remaining total data set, selected in descending order of popularity, so that we now have the 46 most popular (non zero commit) projects per language.

When checking bug-resolution time, we removed projects with no bug-resolution issues (this resulted in 137 projects being excluded).
The language with fewest projects, after this filtering phase, had 35 projects.
Therefore, we chose 35 projects per language (once again, in descending order of popularity).

\pleasecheck{Some issue resolutions consume surprisingly low time (e.g., within one minute). After manual inspection, we found that there are cases which developers may have created issues for already-fixed bugs, then closed them immediately, merely for the purpose of issue recording. 
To ensure that such trivially immediately resolved issues do not adversely affect our results, we removed 3,965 (3.5\%) issues whose resolution time is less than 2 minutes.}

\begin{figure*}[h!]
    \center
      \resizebox{.75\textwidth}{!}{
    \begin{tabular}{ccc}
        \vspace{-3mm}
    \includegraphics[width = 0.3\linewidth,totalheight = 0.19\textheight]{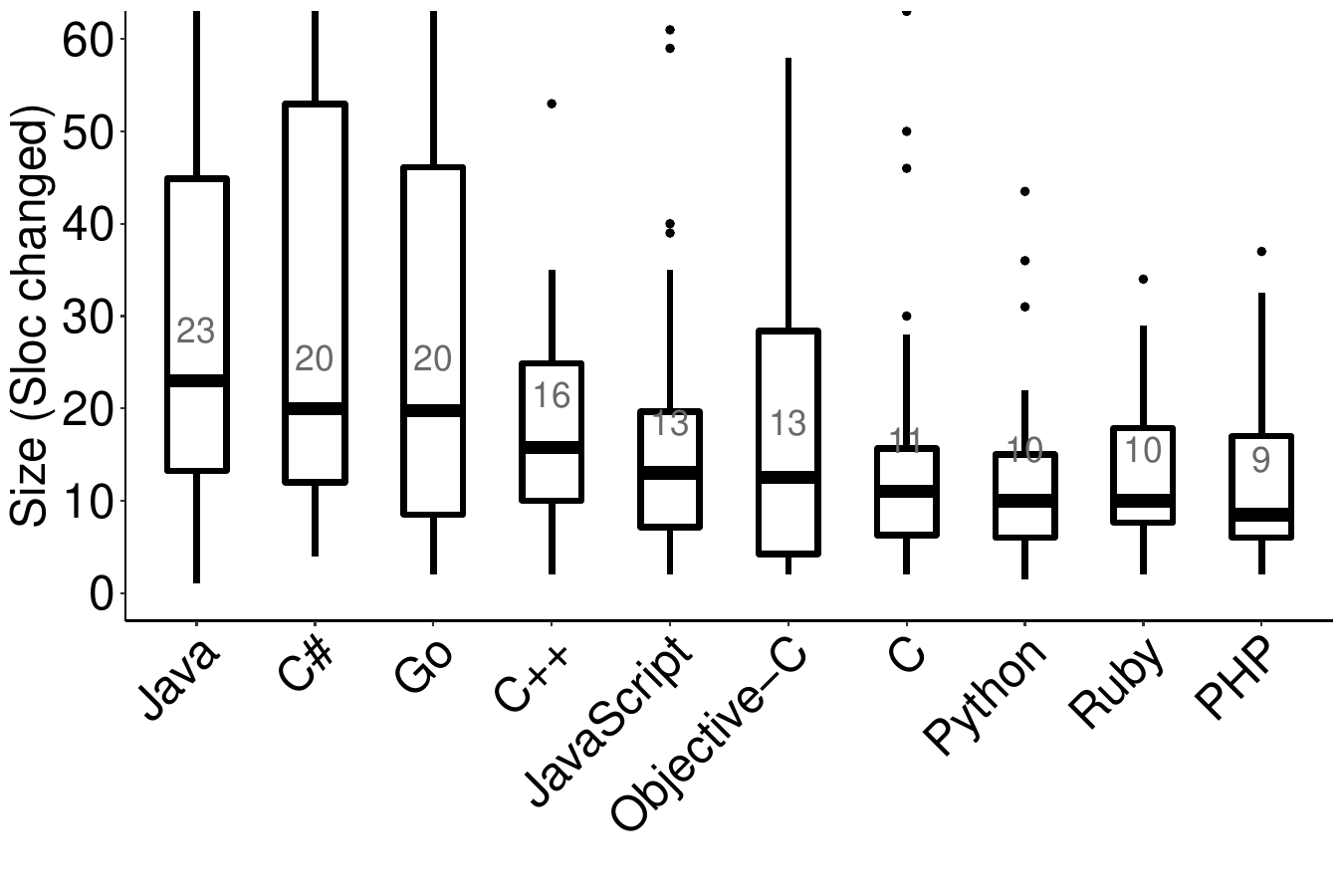}
  \includegraphics[width = 0.3\linewidth,totalheight = 0.19\textheight]{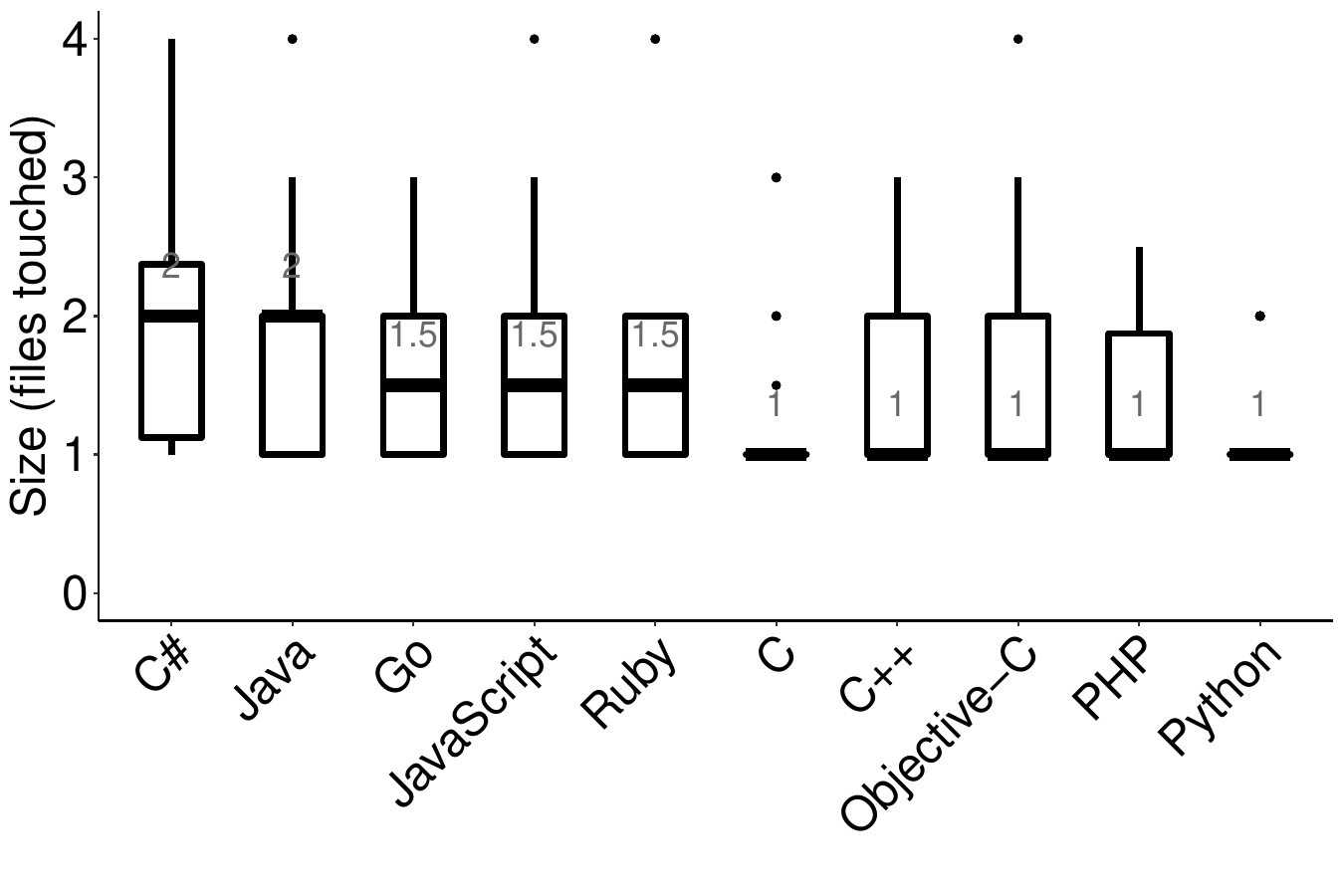}
   \includegraphics[width = 0.3\linewidth,totalheight = 0.19\textheight]{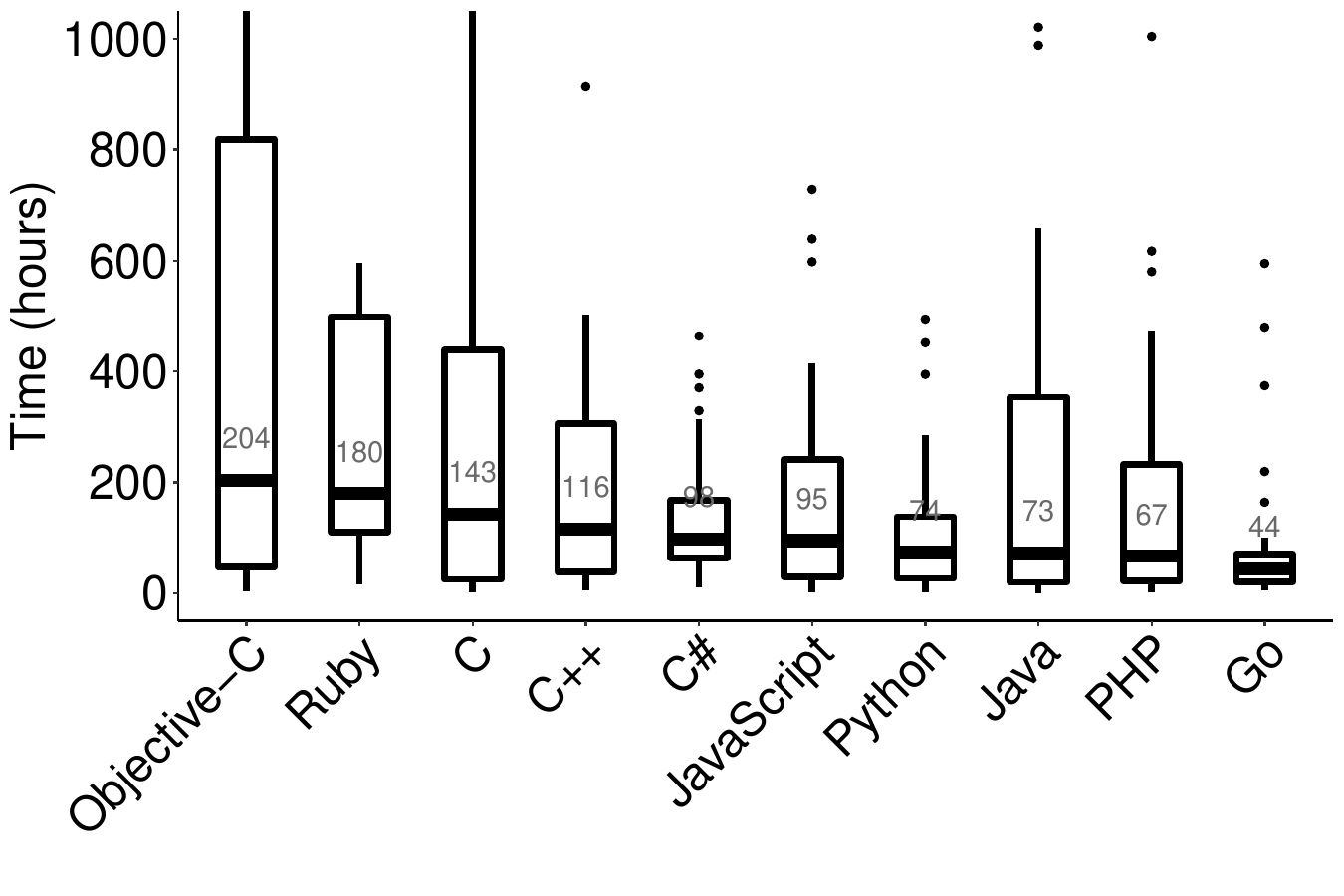}\\
   \includegraphics[width = 0.3\linewidth,totalheight = 0.19\textheight]{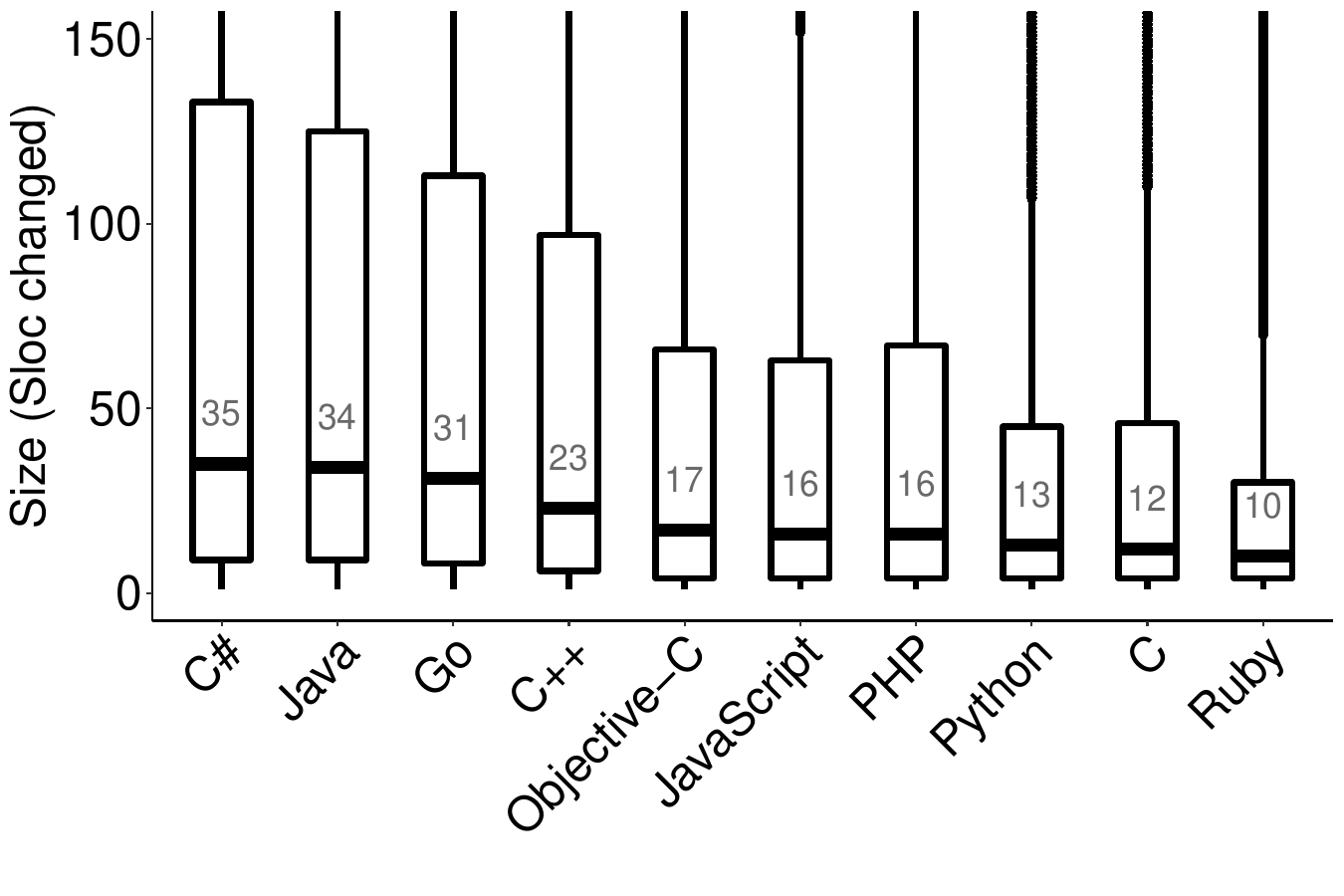}
  \includegraphics[width = 0.3\linewidth,totalheight = 0.19\textheight]{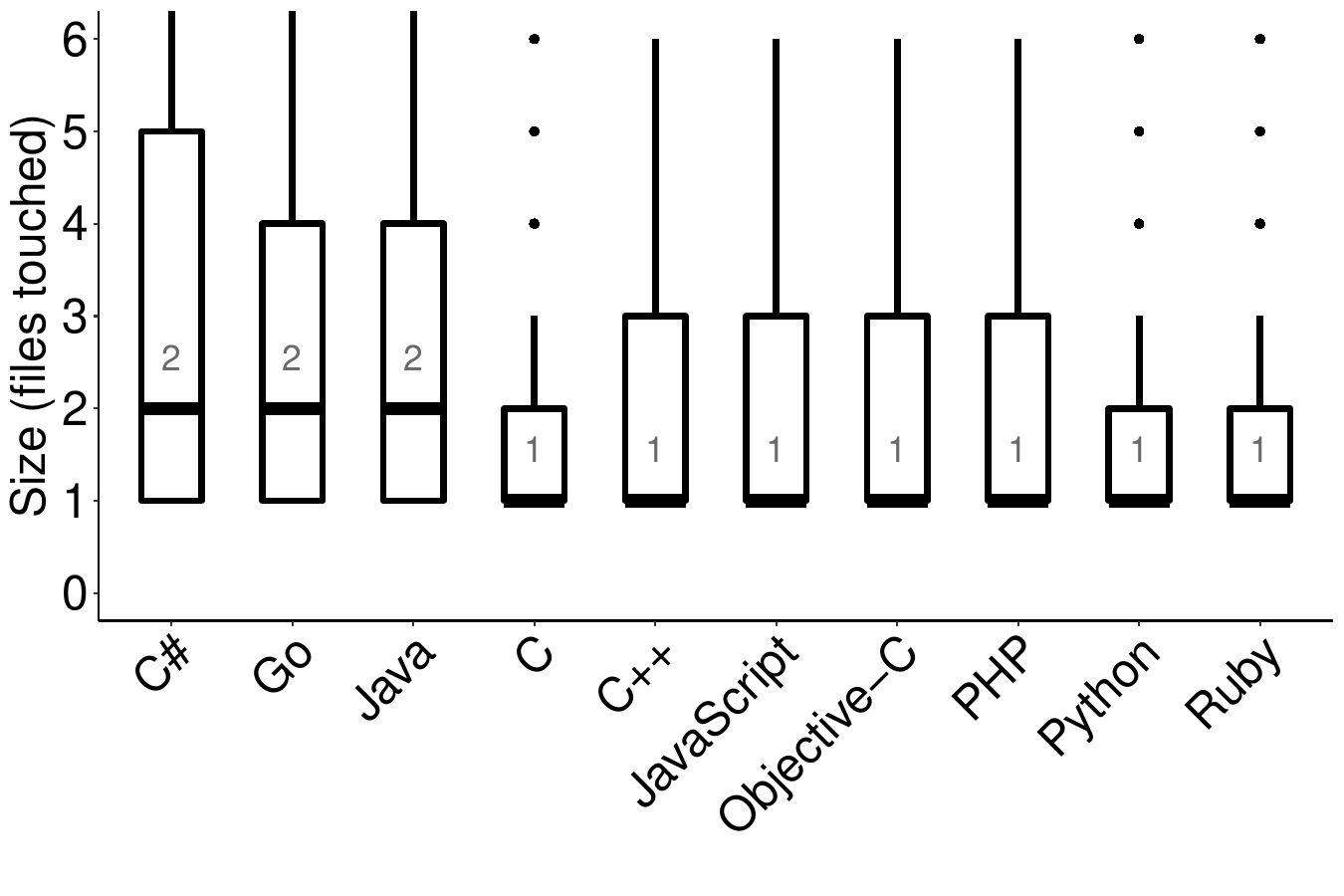}
   \includegraphics[width = 0.3\linewidth,totalheight = 0.19\textheight]{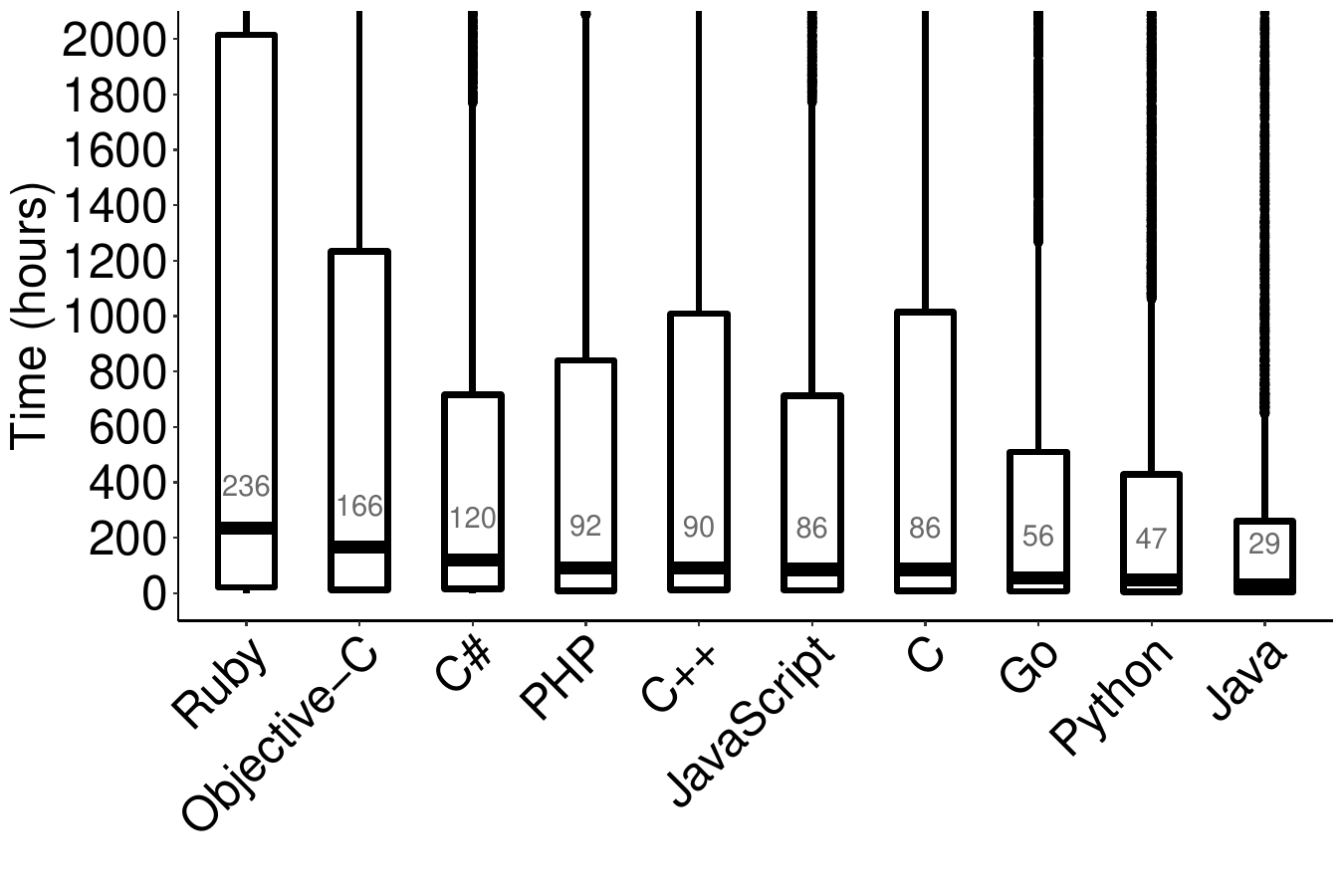}\\
    \end{tabular}}
        \vspace{-1mm}
    \caption{RQ1.1: Median-value comparison. X-axis represents different languages ranked by medians in descending order. The first row is for project-level median (each project contributes a single data point); the second row is for commit-level median (each commit contributes a single data point).}
    \label{fig:median}
\end{figure*}

\subsubsection{Data Analysis}
\label{sec:dataanalysis}

To answer RQ1 and RQ2 we rank the languages and categories based on the median bug-resolution size/time, calculate the multiple regression results (Section~\ref{sec:analysisapproach}), and report the ScottKnott analysis results. 
To answer RQ3, we calculate the correlation between each project feature (including lines of code, project age, number of commits and contributors) and bug-resolution size/time.
We also include these features in multiple regression models and compare their coefficient values with those of the languages.
To answer RQ4, we follow previous work~\cite{ray2014large} by manually classifying projects
into six domains: Application, Database, Codeanalyser, Middleware, Library, and Framework (details are provided in our web page~\cite{thispaperhomepage} for this project, which contains data and results to support replication and re-analysis).
We then analyse the correlation between domains and bug-resolution size/time, to check the extent to which there is evidence for a correlation between application domain and bug-resolution size/time. 

To reduce potential threats to validity arising from  manual classification, two authors classified all the projects independently, and then a third author re-classified the projects where the first two gave conflicting classification. The Cohen's Kappa coefficient~\cite{mchugh2012interrater} 
for inter-rater agreement between the first two raters is 0.734. This indicates a reasonable level of agreement in manual classification outcomes~\cite{Kappa01,Kappa02}, but independent re-analysis and replication would help to further reduce any threats to validity here.

\section{Results and Analysis}
\label{sec:results}
In this section, we present the results of our study. For RQ1 and RQ2, we first give the observations of the three analysis results: median-value analysis, multiple regression analysis, and ScottKnott analysis  (each analysis approach corresponds to one sub-RQ), and then summarise the findings.

\subsection{RQ1: Differences among Programming Languages}
This section introduces our experimental results for answering RQ1.
\subsubsection{RQ1.1: Median Value Comparison}

We rank the bug-resolution size/time \jienew{of programs written in each} language according to the medians of their project efforts in project and commit level analysis respectively.
Figure~\ref{fig:median} shows the results. 
From the figure we observe that, for both types of median value comparison
\java{} and \csharp{} projects exhibit a higher median number of modified lines and files during bug resolution than the other languages, while \python{} and \ruby{} projects \jienew{lie at} the opposite end. 
Additionally, we observe that \java{} have only one eighth (29 vs. 236) of the resolution time of \ruby{} in the commit-level median comparison.

Analysis of the scatter plots provides further insights into the differences observed in the median size of commits, and the median bug resolution time differences between languages.
For example, looking at the scatter plot for bug resolution time for \ruby{} and \java{}, we see that the \java{} projects in our corpus include six with a large number of quickly-resolved issues. 
These kinds of projects consequently have a very low medium resolution time, whereas \ruby{} has only one such project.

\subsubsection{RQ1.2: Multiple Regression} 

Table~\ref{tab:regressionlanguage} shows the regression analysis results. 
The languages are ranked based on their project-level coefficient values (smallest first).

We observe that similar to median-value analysis, projects in the corpus written in \java{} and \csharp{} have the largest 
coefficient values for bug-resolution size for both project-level and commit-level analysis, while those in \python{}, \cc{} and \ruby{} have smallest coefficient values.
For bug-resolution time, projects in \go{} has the smallest coefficient in project-level analysis, projects in \java{} has the smallest coefficient in commit-level analysis.
\ruby{} has the largest coefficients for both project-level and commit-level analysis.

\begin{table}[h!]\small \renewcommand{\arraystretch}{0.6}
    \centering
\caption{RQ1.2\&RQ1.3: Multiple linear regression and ScottKnott results for different languages. In each cell, the first/second value is for project/commit level analysis.}
    \label{tab:regressionlanguage}
    \vspace{0mm}
    \resizebox{.48\textwidth}{!}{
    \begin{tabular}{p{1.7cm}rrr|c}
        \toprule
    \multicolumn{5}{c}{\textbf{Lines of code for bug-resolution} }\\\midrule
      Language&Coeff.&t-value&Pr($>t$)&SK\\ \midrule
\csharp{}&3.34~/~3.62&22.46~/~691.40&$<$2e-16 ~/~$<$2e-16 &a~/~a\\
\java{}&3.21~/~3.57&21.60~/~694.00&$<$2e-16 ~/~$<$2e-16 &a~/~b\\
\go{}&2.93~/~3.49&19.68~/~836.80&$<$2e-16 ~/~$<$2e-16 &a~/~c\\
\cpp{}&2.77~/~3.30&18.62~/~815.50&$<$2e-16 ~/~$<$2e-16 &b~/~d\\
\js{}&2.56~/~3.00&17.20~/~679.50&$<$2e-16 ~/~$<$2e-16 &b~/~e\\
\obc{}&2.51~/~2.99&16.92~/~320.70&$<$2e-16 ~/~$<$2e-16 &b~/~f\\
\ruby{}&2.44~/~2.48&16.39~/~699.50&$<$2e-16 ~/~$<$2e-16 &b~/~h\\
\cc{}&2.37~/~2.72&15.97~/~810.40&$<$2e-16 ~/~$<$2e-16 &b~/~g\\
\php{}&2.29~/~3.00&15.37~/~875.70&$<$2e-16 ~/~$<$2e-16 &b~/~e\\
\python{}&2.25~/~2.72&15.12~/~762.90&$<$2e-16 ~/~$<$2e-16 &b~/~g\\
 \multicolumn{5}{l}{\texttt{F-statistics: 247.4~/~5.384e+05}}\\
 \midrule
  \midrule
\multicolumn{5}{c}{\textbf{Number of files for bug-resolution} }\\\midrule
Language&Coeff.&t-value&Pr($>t$)&SK\\ 
\midrule
\csharp{}&0.77~/~1.03&10.42~/~382.60&$<$2e-16 ~/~$<$2e-16 &a~/~a\\
\java{}&0.55~/~0.94&7.43~/~353.30&5.54e-13~/~$<$2e-16&b~/~b\\
\js{}&0.39~/~0.62&5.36~/~273.40&1.33e-07~/~$<$2e-16&c~/~g\\
\go{}&0.39~/~0.82&5.24~/~381.90&2.51e-07~/~$<$2e-16&c~/~c\\
\ruby{}&0.38~/~0.48&5.16~/~260.30&3.79e-07~/~$<$2e-16&c~/~i\\
\cpp{}&0.35~/~0.71&4.73~/~342.50&3.00e-06~/~$<$2e-16&c~/~e\\
\obc{}&0.33~/~0.70&4.47~/~145.80&9.83e-06~/~$<$2e-16&c~/~f\\
\php{}&0.23~/~0.77&3.09~/~433.60&0.002~/~$<$2e-16&d~/~d\\
\python{}&0.18~/~0.44&2.37~/~238.10&0.018~/~$<$2e-16&d~/~j\\
\cc{}&0.13~/~0.71&1.71~/~342.50&0.088~/~$<$2e-16&d~/~h\\
\multicolumn{5}{l}{\texttt{F-statistics: 30.7~/~1.021e+05}}\\
 \midrule
 \midrule
\multicolumn{5}{c}{\textbf{Time for bug resolution} }\\\midrule
Language&Coeff.&t-value&Pr($>t$)&SK\\       
        \midrule
\ruby{}&5.47~/~5.10&19.09~/~136.91&$<$2e-16~/~$<$2e-16&a~/~a\\
\obc{}&5.19~/~4.68&18.13~/~69.88&$<$2e-16~/~$<$2e-16&a~/~b\\
\cpp{}&4.90~/~4.41&17.11~/~141.03&$<$2e-16~/~$<$2e-16&a~/~c\\
\cc{}&4.70~/~4.39&16.40~/~94.69&$<$2e-16~/~$<$2e-16&a~/~c\\
\csharp{}&4.62~/~4.41&16.14~/~167.58&$<$2e-16 ~/~$<$2e-16 &a~/~c\\
\js{}&4.38~/~4.31&15.28~/~148.63&$<$2e-16 ~/~$<$2e-16 &b~/~c\\
\java{}&4.24~/~3.32&14.81~/~96.78&$<$2e-16 ~/~$<$2e-16 &b~/~f\\
\php{}&4.22~/~4.28&14.74~/~205.19&$<$2e-16 ~/~$<$2e-16 &b~/~c\\
\python{}&4.15~/~3.78&14.49~/~185.64&$<$2e-16 ~/~$<$2e-16 &b~/~e\\
\go{}&3.84~/~3.92&13.40~/~143.28&$<$2e-16 ~/~$<$2e-16 &b~/~d\\
\multicolumn{5}{l}{\texttt{F-statistics: 257.5~/~2.091e+04}}\\
        \bottomrule
    \end{tabular}
    }
    \vspace{-2mm}
\end{table}

\subsubsection{RQ1.3: ScottKnott Analysis}

\pleasecheck{
The results of ScottKnott analysis are shown by Column `SK' of Table~\ref{tab:regressionlanguage}. 
Each letter represents one clustered group. 
The ranking of letters corresponds to the ranking of languages based on treatment means. 
}

\pleasecheck{
We observe that each analysis has at least two groups, indicating that significant differences exist among the projects written in different programming languages in our corpus for bug-resolution size/time. 
Commit-level analysis yields more groups than project-level analysis.
This is because in project-level analysis, there is one data point for each project with a median bug-resolution size/time among all commits of this project, and the diversity of bug-resolution size/time is reduced. }

\pleasecheck{
For commit-level analysis, the rankings of different programming languages are consistent with the results we witnessed using both multiple regression and median-value comparison. We do not repeat the observations here, but summarise our conclusions from the three analyses in the below.}

\begin{tcolorbox}
\textbf{Finding for RQ1}:  
\jienew{In our corpus, projects written in} \java{} and \csharp{} exhibit notably higher bug-resolution size than the other languages.   
\jienew{Projects written in} \ruby{} consume notably higher bug-resolution time than \jienew{those written in} the other languages. 
\end{tcolorbox}

\pleasecheck{In particular, if we use the median bug-resolution size/time of the projects in each language to conduct quantitative comparison, we have the following  findings for the programs in our corpus that, we believe, merit closer scrutiny, re-analysis and replication attempts in future work: 1) \java{} and \csharp{} bug-resolutions occupy 2/3 times as many lines as \ruby{} and \python{} in project/commit-level analysis; 2) \java{} and \csharp{} bug-resolutions touch 2 times as many files as \ruby{} and \python{}; 3) \ruby{} bug-resolution consumes 2.5/8.1 times as much time as \java{} in project/commit-level analysis.}

\subsubsection*{Connection between RQ1 and Great Debates}
In the following, we examine the results within each language, exploring the extent to which they confirm and relate to existing perceptions of the language.
Clearly with such a large scale study (with so many potentially confounding factors), any analysis based on our initial results will be somewhat speculative. 
Nevertheless, we include this discussion here, and its relation to great debates on programming languages, as one way of highlighting potential avenues for future work.

An interesting multi language pattern we can observe is that \go{} and \java{} both have more line/file modifications yet exhibit lower bug-resolution time, which leads to the conclusion that \emph{programs occupying more line/file modifications do not necessarily consume more bug-resolution time.}
\pleasecheck{Similarly, \ruby{} has fewer line/file modifications yet exhibits longer bug-resolution time, which leads to the conclusion that \emph{programs occupying fewer line/file modifications do not necessarily consume less bug-resolution time.}
}
If replicated, more widely, this pattern may partially explain the great debates regarding the perceived impact of programming languages on bug-resolution characteristics.

We now turn to exploration of the results for each individual language, and the connection between these and  claims in the literature about such languages.

{\bf Results on \java{}.}
\java{} project bug fixes \emph{occupy more line/file modification, yet consume less bug-resolution time.} The finding about bug-resolution size concurs with the claim that \java{} is a verbose language~\cite{broussard2006method}. Our results indicate that this perceived verbosity may carry over to bug resolution (and, indeed, to commit size more generally, see Section~\ref{sec:non-buggy}). 

\jienew{Despite the perceived verbosity we find}, \java{} is one of the languages with lower bug-resolution time. 
In particular, in commit-level analysis, \java{} projects have the least bug-resolution time.
One reason may relate to the claim that there are large number of declarations required in \java{}, such as type declaration, method parameter types, return types, access levels of classes, exception handling~\cite{broussard2006method}.
\jienew{Perhaps this} makes the language verbose yet at the same time provides additional documentation, making the code easier to understand and, thereby, to debug~\cite{arnold2000java}.

{\bf Results on \python{}.}  \python{} \jienew{projects in our corpus} \emph{occupy less bug-resolution size and time.} 
\python{} is widely believed to have a large set of  libraries and a very active community, which may make it easier for developers to find support during bug resolution. It is also reported that there has been a trend in the \python{} community to improve code quality by dictating ``one right way''~\cite{startup}. The maturity of the community and the effort of adhering to best practices may facilitate bug resolution.

{\bf Results on \ruby{}.} 
\ruby{} \jienew{projects in our corpus exhibit} \emph{longer bug-resolution time. } As a dynamic language, \ruby{} is designed to make programming a ``pleasant" and ``productive" experience~\cite{rubyvspython}, which does not have hard rules on writing code~\cite{flanagan2008ruby}.
One example of the flexible features of \ruby{} is ``monkey patching'', which refers to the extension or modification of existing code by changing classes at run-time, where
any class can be re-opened at any time and amended in any way. It is a flexible technique that has become popular in the \ruby{} community; but this flexibility may lead to hard-to-diagnose clashes~\cite{monkeypatching}.

\jie{We investigated how frequently ``monkey patching'' is used within the Ruby projects we studied.
As ``monkey patching'' refers to dynamic modifications of a class or module at runtime, it is not easy to judge whether a project has adopted it merely based on static analysis. 
Instead, we searched all Ruby project sources (comment, commit messages, issue discussions) using keyword ``monkey patch''. We then manually confirmed the search hits. 
Overall, key word searching helped us to confirm that at least 40 out of the original 57 Ruby projects have used monkey patching.}

{\bf Results on \cc{}, \cpp{} and \obc{}.} When comparing \cc{}, \cpp{} and \obc{}, we observe that projects in
\obc{} exhibit higher bug-resolution time. We note that \obc{} has a mixture of static and dynamic typing, whereas plain \cc{} and \cpp{} objects are statically typed.

To investigate, in more detail, how often dynamic typing is used in the \obc{}  programs we studied, we manually analysed the source code files of the \obc{} projects. As mentioned in the Apple \jienew{documentation}\footnote{\url{https://developer.apple.com/library/archive/documentation/General/Conceptual/DevPedia-CocoaCore/DynamicTyping.html}}, Objective-C uses the \texttt{id} data type to represent a variable that is an object without specifying what sort of object it is (dynamic typing). \jienew{With} this information \jienew{and manual confirmation}, we found that 90\% of our projects have adopted dynamic typing.

\subsection{RQ2: Differences among Language Categories}

To answer the second research question, we apply the three analysis approaches to the same projects as we used in RQ1. 
In RQ1, each project was labelled with a single programming language, whereas in RQ2, each project has two labels according to its language's approach to types: dynamically or statically typed, and strongly or weakly typed. 
We first compare dynamically and statically typed \jienew{sets of programs}, then compare strongly and weakly typed \jienew{sets of programs}.

Naturally, such categorisations are fraught with potential issues, so we must be careful in drawing conclusions.
Nevertheless, where \pleasecheck{differences in median values} observed are sufficiently large, this may highlight potential avenues for which follow-on research, re-analysis and replication effort are worthwhile.

\subsubsection{RQ2.1: Median-value analysis}

\begin{figure}[t]
    \center
     \resizebox{.45\textwidth}{!}{
    \begin{tabular}{cc}
        \vspace{-3mm}
    \includegraphics[width = 0.44\linewidth,totalheight = 0.11\textheight]{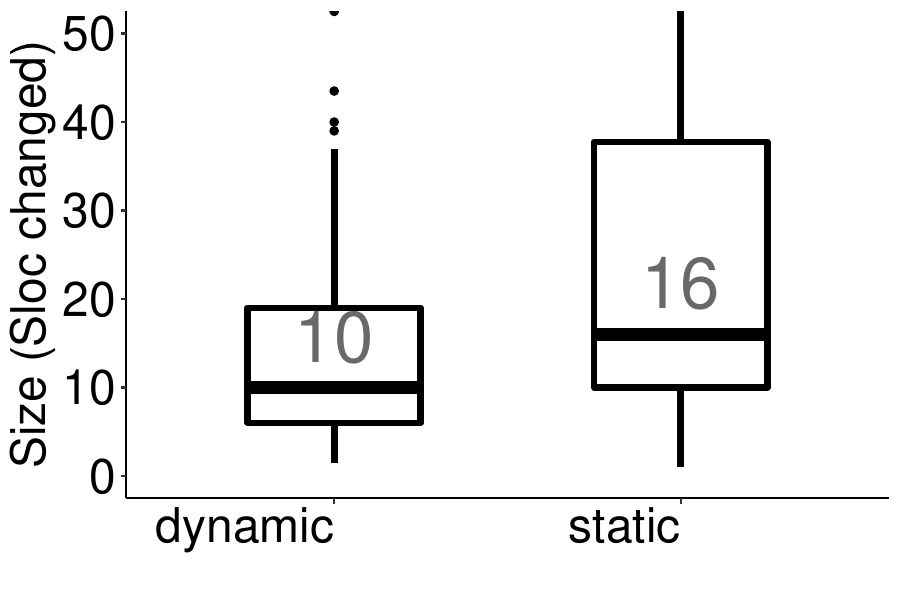}
    \hspace{3mm}\includegraphics[width = 0.44\linewidth,totalheight = 0.11\textheight]{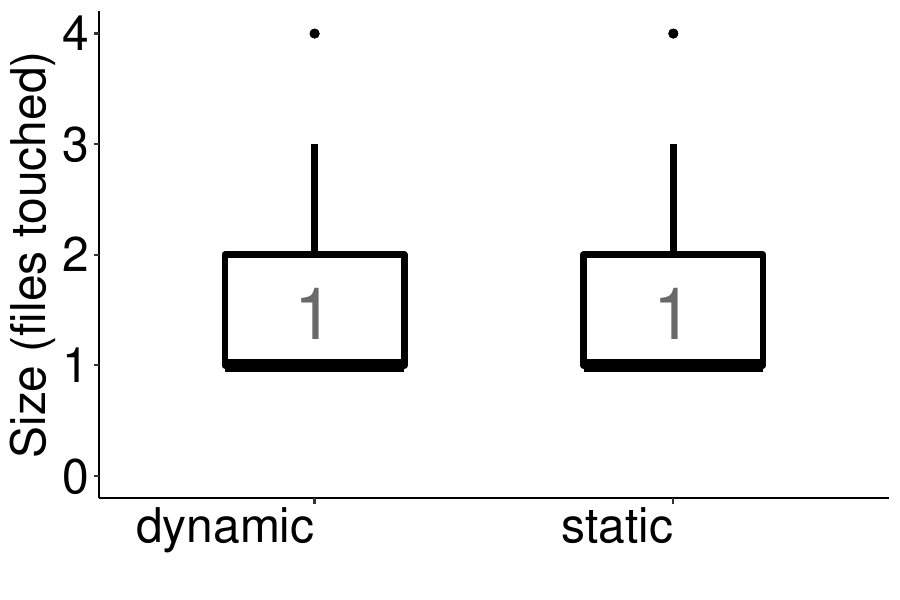}
    \hspace{3mm}\includegraphics[width = 0.44\linewidth,totalheight = 0.11\textheight]{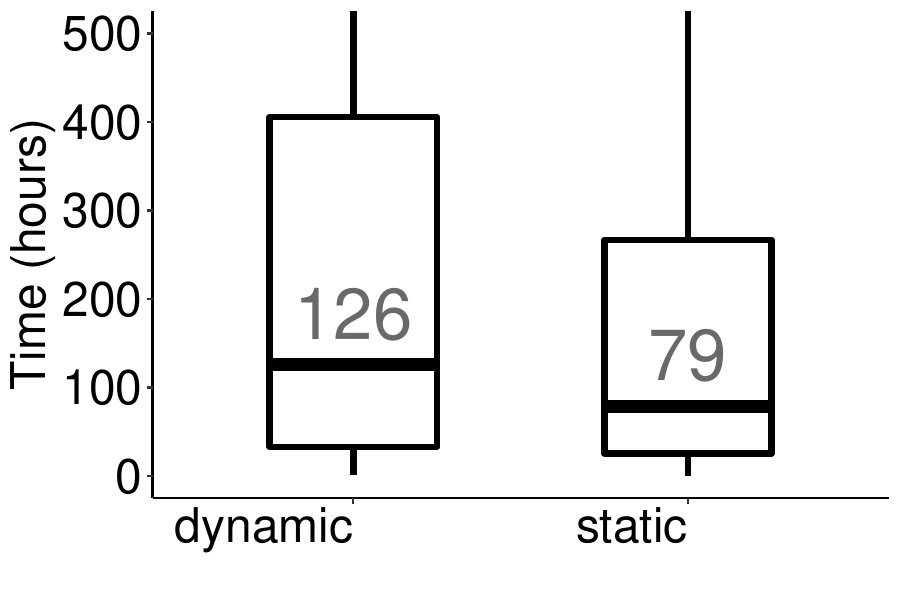}\\
   
    \includegraphics[width = 0.44\linewidth,totalheight = 0.11\textheight]{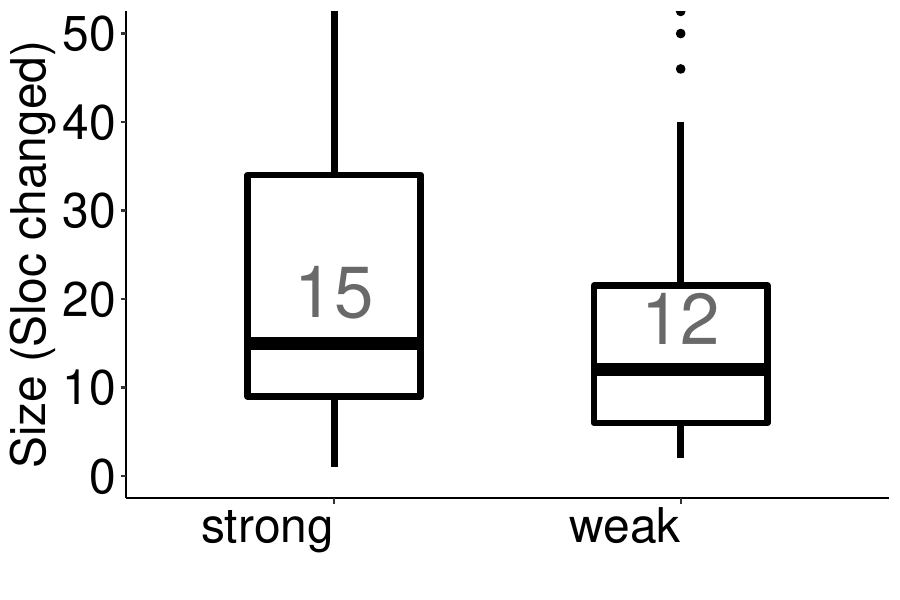}
    \hspace{3mm}\includegraphics[width = 0.44\linewidth,totalheight = 0.11\textheight]{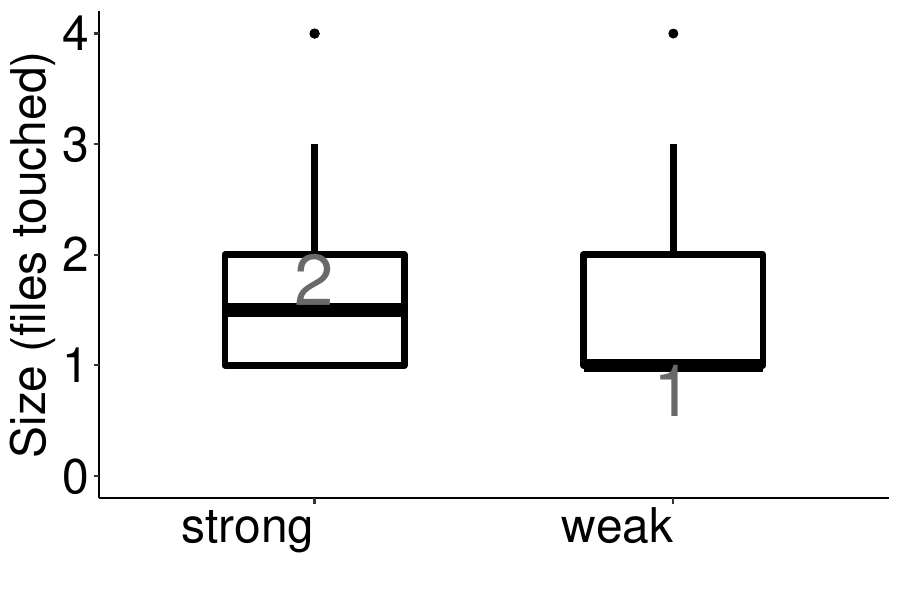}
    \hspace{3mm}\includegraphics[width = 0.44\linewidth,totalheight = 0.11\textheight]{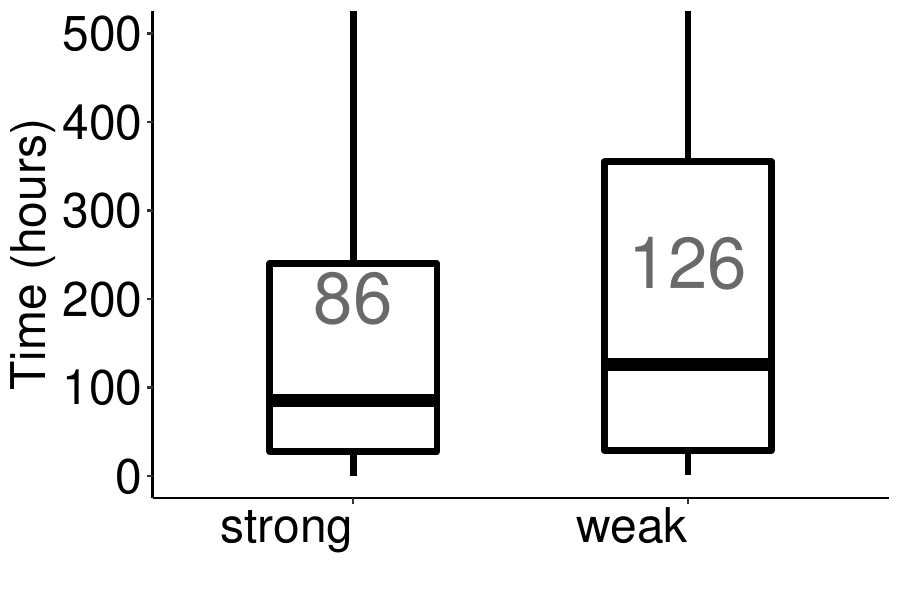}\\
     \includegraphics[width = 0.44\linewidth,totalheight = 0.11\textheight]{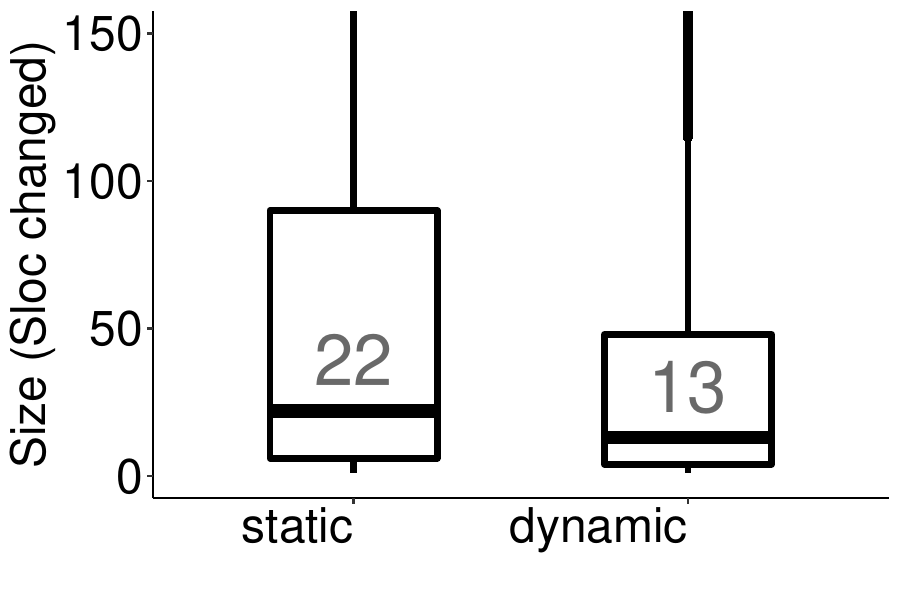}
    \hspace{3mm}\includegraphics[width = 0.44\linewidth,totalheight = 0.11\textheight]{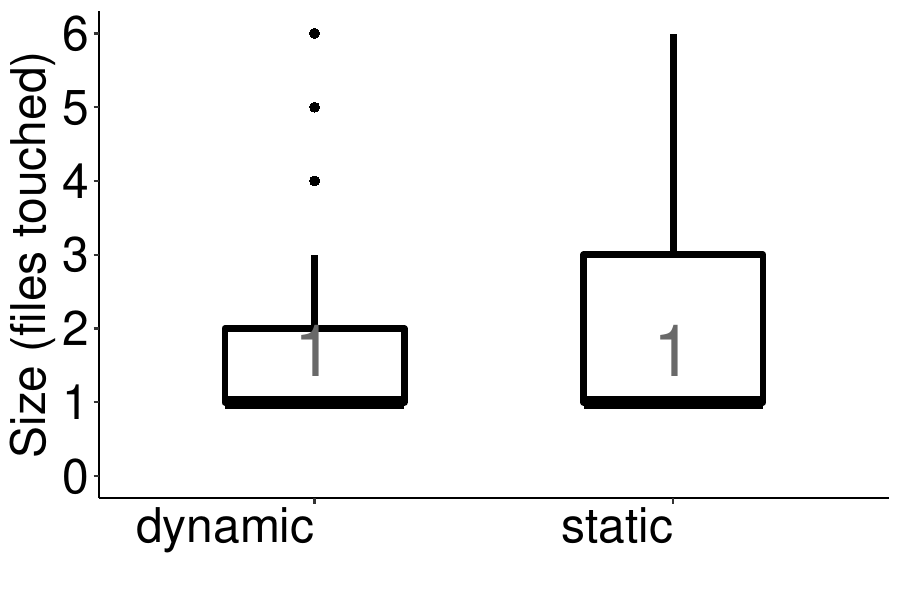}
    \hspace{3mm}\includegraphics[width = 0.44\linewidth,totalheight = 0.11\textheight]{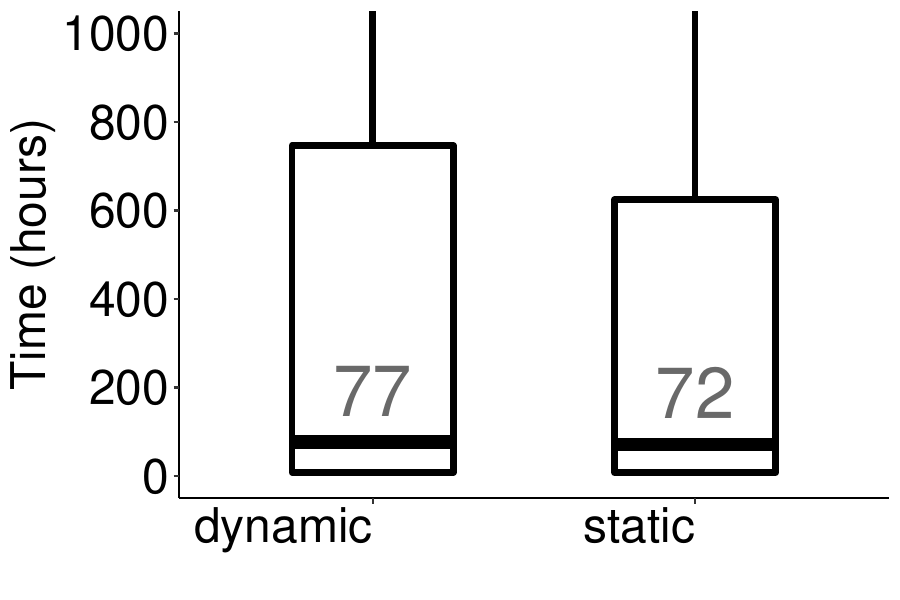}\\
    \includegraphics[width = 0.44\linewidth,totalheight = 0.11\textheight]{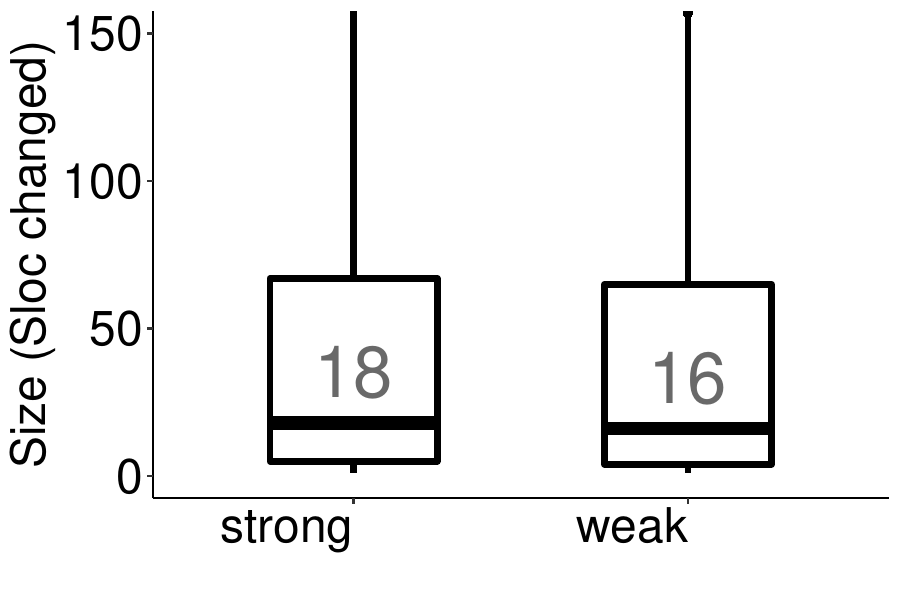}
    \hspace{3mm}\includegraphics[width = 0.44\linewidth,totalheight = 0.11\textheight]{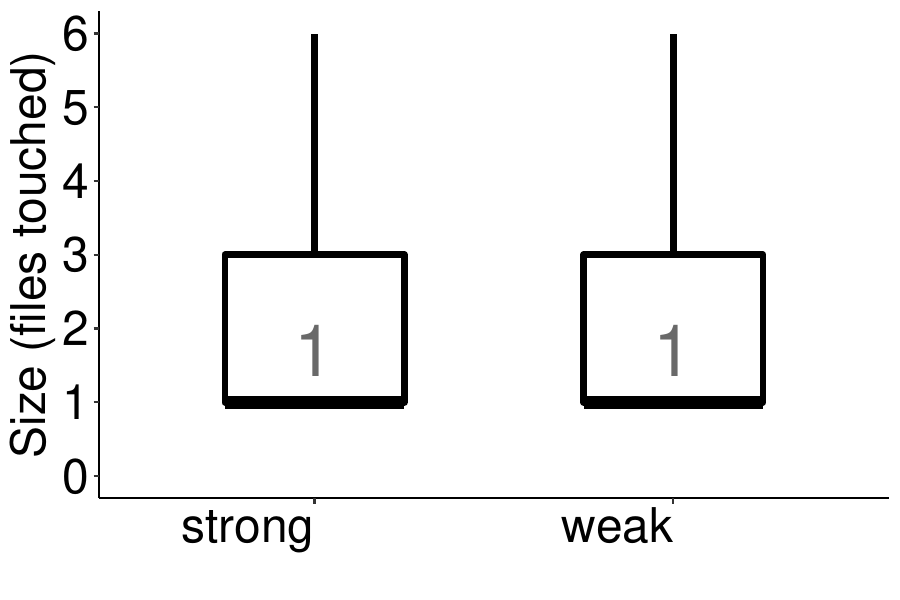}
    \hspace{3mm}\includegraphics[width = 0.44\linewidth,totalheight = 0.11\textheight]{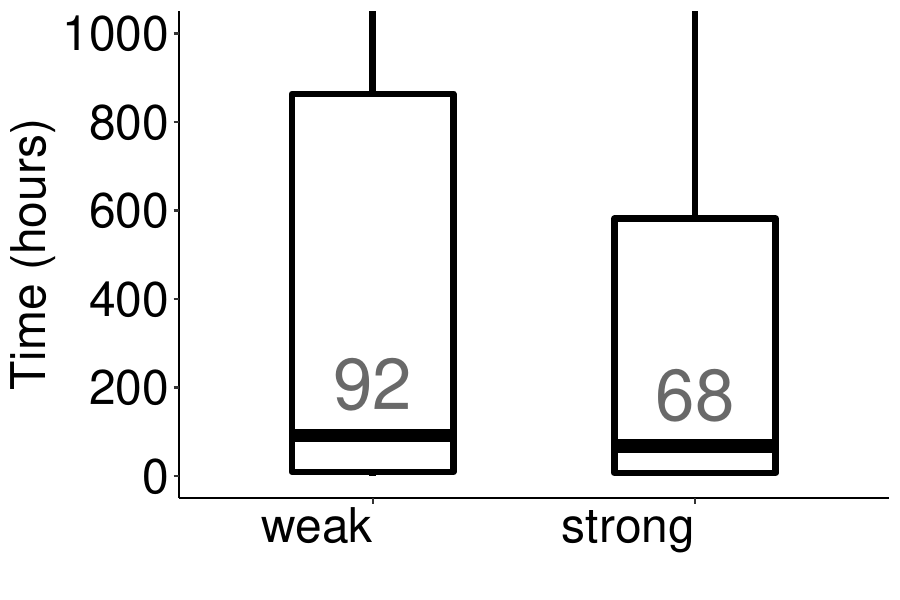}\\
    \end{tabular}}
        \vspace{-1mm}
    \caption{RQ2.2: Median value comparison. The first/second two rows are for project/commit-level analysis. }
    \label{fig:mediancategory}
\end{figure}

We give the results of the median-value analysis as shown in Figure~\ref{fig:mediancategory}. 
From these box plots, differences are revealed between different language categories for \ploc{} and \ptime{}. 
In particular, dynamically/weakly typed program sets occupy fewer line modifications, but longer bug-resolution time.

\subsubsection{RQ2.2: Multiple Regression}

The same as RQ1.2, we perform multiple linear regression analysis, and observe the coefficient values of different language categories. 
Larger coefficients indicate more contributions the variable has to the regressed model.
Table~\ref{tab:categorycombination} shows the results. 
In the table, we observe that dynamically typed program sets have lower coefficient values in bug-resolution size but a higher coefficient value in bug-resolution time, than statically typed program sets. 
The p values and t values give some degree of confidence in these observations, but further study with other corpora and re-analysis would be beneficial to confirm or refute.

Based on these observations,  there is  evidence  that dynamically typed program sets contribute less to bug-resolution size but more to bug-resolution time than statically typed \jienew{program sets} in the regression model.
Similarly, we observe that bug fixes in weakly typed \jienew{program sets}  occupy more lines and files than strongly typed \jienew{program sets}, but  consume lower bug-resolution time.

\begin{table}[t]\small
\centering
\caption{RQ2.2\&RQ2.3: Multiple regression and ScottKnott results for different language categories. }
\label{tab:categorycombination}
\vspace{0mm}
  \resizebox{.48\textwidth}{!}{
    \begin{tabular}{p{1.7cm}rrr|c}
\toprule
    \multicolumn{5}{c}{\textbf{Lines of code for bug-resolution} }\\\midrule
      Category&Coeff.&t-value&Pr($>t$)&SK\\ \midrule
static{}&2.92~/~3.23&42.97~/~1694&$<$2e-16 ~/~$<$2e-16 &a~/~a\\
dynamic{}&2.41~/~2.79&35.40~/~1525&$<$2e-16 ~/~$<$2e-16 &b~/~b\\
\multicolumn{5}{l}{\texttt{F-statistics: 1550~/~2.597e+06}}\\
\midrule
strong{}&2.83~/~3.03&40.88~/~1599&$<$2e-16 ~/~$<$2e-16 &a~/~a\\
weak{}&2.50~/~2.98&36.09~/~1593&$<$2e-16 ~/~$<$2e-16 &b~/~b\\
\multicolumn{5}{l}{\texttt{F-statistics: 1487~/~2.547e+06}}\\
 \midrule
 \midrule
\multicolumn{5}{c}{\textbf{Number of files for bug-resolution} }\\\midrule
Category&Coeff.&t-value&Pr($>t$)&SK\\ 
\midrule
static{}&0.43~/~0.74&12.65~/~745.10&$<$2e-16 ~/~$<$2e-16 &a~/~a\\
dynamic{}&0.30~/~0.58&8.76~/~612.50&$<$2e-16 ~/~$<$2e-16 &b~/~b\\ 
\multicolumn{5}{l}{\texttt{F-statistics: 118.4 ~/~4.651e+05}}\\
\midrule
strong{}&0.45~/~0.66&13.18~/~680.60&$<$2e-16 ~/~$<$2e-16 &a~/~a\\
weak{}&0.28~/~0.64&8.33~/~669.50&9.34e-16~/~$<$2e-16&b~/~b\\
\multicolumn{5}{l}{\texttt{F-statistics: 121.5~/~4.557e+05}}\\
 \midrule
 \midrule
\multicolumn{5}{c}{\textbf{Time for bug resolution} }\\\midrule
Category&Coeff.&t-value&Pr($>t$)&SK\\       
        \midrule
dynamic{}&4.68~/~4.21&35.61~/~345.20&$<$2e-16 ~/~$<$2e-16 &a~/~a\\
static{}&4.46~/~4.10&33.93~/~290.70&$<$2e-16 ~/~$<$2e-16 &a~/~b\\
\multicolumn{5}{l}{\texttt{F-statistics: 1210~/~1.018e+05}}\\
\midrule
weak&4.68~/~4.34&35.58~/~310.50&$<$2e-16 ~/~$<$2e-16 &a~/~a\\
strong&4.47~/~4.03&33.96~/~328.50&$<$2e-16 ~/~$<$2e-16 &a~/~b\\
\multicolumn{5}{l}{\texttt{F-statistics: 1210~/~1.022e+05}}\\
        \bottomrule
    \end{tabular}
    }
    \vspace{-2mm}
\end{table}

\subsubsection{RQ2.3: ScottKnott analysis}
\pleasecheck{
To answer RQ2.3, we report the results of the ScottKnott analysis in Column `SK' of Table~\ref{tab:categorycombination}. 
For bug-resolution size, different categories are clustered into different groups.
This means both project-level and commit-level analysis reveal significant differences (significance level: 0.05) between different categories.
For bug-resolution time, there are significant difference for commit-level analysis.
Nevertheless, for project-level analysis, we did not observe such significance. 

Compared with median-value comparison in RQ2.1 and multiple regression in RQ2.2, 
statically/strongly typed program sets tend to occupy more line/file modifications yet lower bug-resolution time compared to dynamically/weakly typed program sets, but the difference is not large for bug-resolution time. 
}

As mentioned in Section~\ref{sec:categories}, four \python{} projects in our dataset contain a small amount of static typing. 
We checked the number of files containing static typing for these four projects and found that projects with more static typing tend to consume less bug-resolution time. Obviously, this is not a finding that we can support with inferential statistical analysis due to the small number of projects with static typing. Nevertheless, it is an interesting observation and suggests further research to compare static and dynamic typing within the same language.

The community migration from an untyped language to a typed language will make an excellent opportunity for more in-depth study.
Perhaps this might be an avenue for further research, since project communities that migrate, while retaining the same ecosystem and developer teams will present an opportunity to study what happens when projects become typed.
Also, there may be fewer confounding factors, when such ecosystems remain in place throughout the migration.

Overall our results point to the following finding:
\begin{tcolorbox}
\textbf{Finding for RQ2}: Strongly/statically typed \jienew{programs in our corpus have} bug resolutions that tend to occupy  more lines and files. 
Dynamically typed programs tend to consume longer bug-resolution time.
\end{tcolorbox}

\pleasecheck{In particular, if we use the median bug-resolution size/time of the projects in each language category to conduct quantitative comparison, we have the following findings for the programs in our corpus: 1) dynamically typed language bug-resolutions occupy 37.5\% fewer lines, yet consumes 60.6\% more bug-resolution time than statically typed languages; 
2) weakly typed language bug-resolutions occupy 20\% fewer lines, yet consume 47.2\% more bug-resolution time than strongly typed languages.}

We calculate time and size of a bug resolution in different ways and one different (but overlapping) set of projects, so these observations are not directly pairwise comparable.
Nevertheless, the observation of these differences over the corpora studied does seem intriguing and, therefore, perhaps a worthy priority for follow-on study.

\pleasecheck{
\subsubsection*{Connection between RQ2 and Great Debates}
As with RQ1, we observe that statically and strongly typed projects have more line/file modifications yet exhibit lower bug-resolution time, which again leads to the conclusion that projects occupying more line/file modifications do not necessarily consume more bug-resolution time.

This observation may partially explain the great debates regarding the perceived impact of types on bug-resolution characteristics.
In particular, programmers or researchers may have used different measurement criteria, e.g., the amount of line modification or the amount of time spent in bug resolution, and consequently, might have drawn apparently contradictory conclusions. 

We see some evidence for this in the narratives found in the literature hitherto. 
For example, Kleinschmager et al.~\cite{kleinschmager2012static} and Hanenberg et al.~\cite{hanenberg2014empirical} presented results that suggest that statically typed languages have lower bug-resolution time. However, their empirical studies used bug-resolution time as the sole measurement criterion.
By contrast, Tratt et al.~\cite{tratt2009dynamically} called statically typed languages ``the enemy of change'' because, they claim, statically typed languages require more complex code modifications.
Our results indicate that both claims may be correct, yet not necessarily inconsistent.

At the same time, we observe that statically and dynamically typed commits have similar median bug-resolution time in our commit-level analysis.
This may provide another explanation for the existence of the great debates over types in languages.
}

\subsection{RQ3: Correlation between bug resolution characteristics and other Project Features}
\label{sec:rq3}

To answer RQ3, we calculate the Pearson/Kendall's $\tau$/Spearman correlation between the four project features and bug-resolution size/time.  %
\pleasecheck{
Table~\ref{tab:featurecorrelation} presents the correlation results. 
For each feature, the first row shows the correlation coefficient values, the second row shows the corresponding p-values.
From the table, most correlation coefficients are below 0.15, 
indicating that the project features we studied have very weak or no correlation with bug-resolution size and time. }

\begin{table}[h!]\small
\centering
\caption{RQ3: Correlation (Pearson/Kendall's $\tau$/Spearman)between project features and bug-resolution characteristics. The second row for each feature shows p-values for each correlation analysis.}
\label{tab:featurecorrelation}
\resizebox{.49\textwidth}{!}{
\begin{tabular}{lrrr}
\toprule
Feature&Size (Sloc changed)&Size (files touched)&Time (hours)\\
\midrule
sloc&-0.01/0.16/0.24&0.00/0.09/0.12&-0.07/-0.06/-0.09\\
&\texttt{0.85/2e-07/3e-07}&\texttt{0.96/0.01/0.46}&\texttt{0.18/0.10/0.08}\\
commit&-0.03/0.05/0.08&-0.03/0.03/0.04&-0.11/ -0.05/-0.07\\
&\texttt{0.54/0.08/0.09}&\texttt{0.46/0.44/0.43}&\texttt{0.04/0.14/0.16}\\
age&-0.11/-0.09/-0.14&-0.12/-0.04/-0.05&0.12/0.18/0.26\\
&\texttt{0.02/0.00/0.00}&\texttt{0.01/0.33/0.33}&\texttt{0.02/7e-07/6e-07}\\
contributor&-0.06/-0.02/ -0.03&-0.07/ -0.01/ -0.01&-0.11/0.00/0.01\\
&\texttt{0.03/0.56/0.58}&\texttt{0.12/0.80/0.77}&\texttt{0.05/0.99/0.89}\\
\bottomrule
\end{tabular}}
\end{table}
\pleasecheck{
In addition, we investigate the impact of project features on bug-resolution characteristics by including them in the multiple regression model. 
We then compare the coefficients of the project features with different languages, to see whether the regression is dominated by any project features.

The results are shown by Table~\ref{tab:regressionwithfeatures}.
For ease of comparison between project feature coefficients and language coefficients, we present the the smallest coefficient for languages in the last row.
From this table, we observe that the coefficients of project features are all smaller than language coefficients.
}

\begin{table}[h!]\small
\centering
\caption{RQ3: Coefficients of project features in multiple regression. }
\label{tab:regressionwithfeatures}
\resizebox{.49\textwidth}{!}{
\begin{tabular}{lrrr}
\toprule
Feature&Size (Sloc changed)&Size (files touched)&Time (hours)\\
\midrule
sloc&0.15&0.73& 0.05\\
commit&-0.07&-1.55&-0.36\\
age&-0.21&-1.81&0.82\\
contributor&-0.00&0.61& 0.24\\
language (smallest)&1.54&4.56&4.05\\
\bottomrule
\end{tabular}}
\end{table}

These observations tend to suggest that the differences we observed among different languages and categories are less likely to be associated with project LOC, age, the number of commits, or the number of contributors.

\vspace{0mm}
\begin{tcolorbox}
\textbf{Finding for RQ3}: We found little evidence that project SLOC, commit number, project age, and contributor number have a strong correlation with bug-resolution size or time.
\end{tcolorbox}

\subsection{RQ4:  Correlation between bug resolution characteristics and application domain}
\label{sec:domainimpact}

We have shown that bug-resolution characteristics are different among different categories of programs. However, since \jienew{programming languages may be designed} with a specific application domain in mind, we would like to know whether the above findings are domain dependent. 
In this section, we address this question by examining the correlation between domain and bug-resolution size/time.

As a start, we count the number of projects belonging to each domain for each language (details in Section~\ref{sec:dataanalysis}) for the 460 projects adopted in bug-resolution size analysis.
We then check whether there is a strong connection between languages and domains. 
Table~\ref{tab:domainmatrix} shows the results. The numbers in the table represent the number of projects \jienew{written primarily in} the given language and domain. For example, there are 14 projects in \cc{} that belong to the \emph{Application} domain.
From the rows of the table, we observe that 
all languages have a diverse domain distribution, with each containing at least five domains.
From the columns of the table, we can see that 
the number of projects is similar across different languages, with the \emph{Library} domain (shown by Column `Lib.') having more projects for most languages.

\jie{
We also calculated the Cramer's V value between domains and languages, which is another correlation-based measure of association between two nominal (categorical in our case) variables~\cite{cramer1999mathematical}. 
The value of this statistic is merely 0.20, indicating a weak association betweenthe two variables.

These observations indicate that, for our subjects, there is no strong correlation between languages and domains.
We thus have the following conclusion:
}

\begin{table}[t]\small
\centering
\caption{Projects of different domains inside each language.}
\label{tab:domainmatrix}
\resizebox{.48\textwidth}{!}{
\begin{tabular}{lrrrrrrrr}
\toprule
Language&App.&Database&CA&MW&Lib.&Frame.&Others\\
\midrule
C&14&3&4&4&12&5&4\\
\csharp{}&10&3&0&16&13&4&0\\
\cpp{}&9&3&5&6&18&2&3\\
Go&14&2&9&9&5&5&2\\
Java&7&2&2&8&23&1&3\\
JavaScript&3&2&0&17&20&1&3\\
Objective-C&3&4&3&11&24&1&0\\
PHP&5&5&0&18&12&2&3\\
Python&11&3&1&11&7&6&7\\
Ruby&8&6&0&13&16&1&2\\
\bottomrule
\end{tabular}}
\end{table}

\begin{tcolorbox}
\textbf{Finding for RQ4}: We found little evidence to suggest that any application domain we explored was  strongly correlated to bug resolution size/time.
\end{tcolorbox}

\section{Extended Analysis}
\label{sec:extendedanalysis}
In this section, we extend the analysis by looking
at a wider set of characteristics of different languages
and their connections to bug-resolution.

\subsection{Aggregated and Normalised Bug-resolution Size}
\label{sec:aggregatedresolutionsize}

The previous sections evaluate bug-resolution size from two aspects: the number of modified lines  and files.
These two aspects are studied separately.
Developers and researchers may be also interested in the aggregated  measurement criteria line*file (i.e., the product effect of modified lines and files), or the normalised measurement criteria 
line/file (i.e., how many lines are modified per file on average in bug resolution).
We present the results for these two measurement criteria in this section.

\begin{figure}[h]
    \center
     \resizebox{.35\textwidth}{!}{
    \begin{tabular}{c}
        \vspace{-3mm}
    \includegraphics[width = 0.76\linewidth,totalheight = 0.18\textheight]{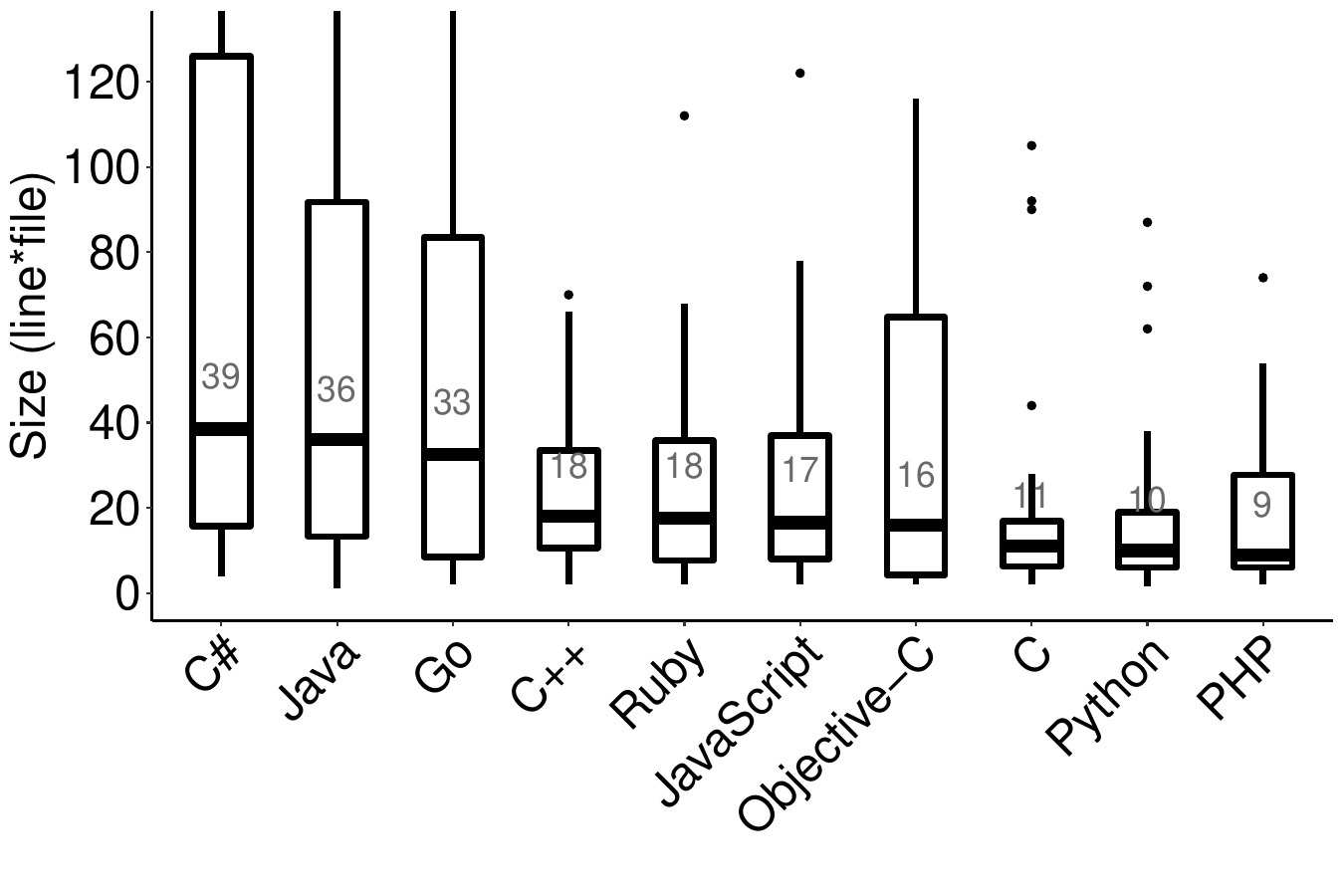}
     \\
  \hspace{1.5mm}   \includegraphics[width = 0.76\linewidth,totalheight = 0.18\textheight]{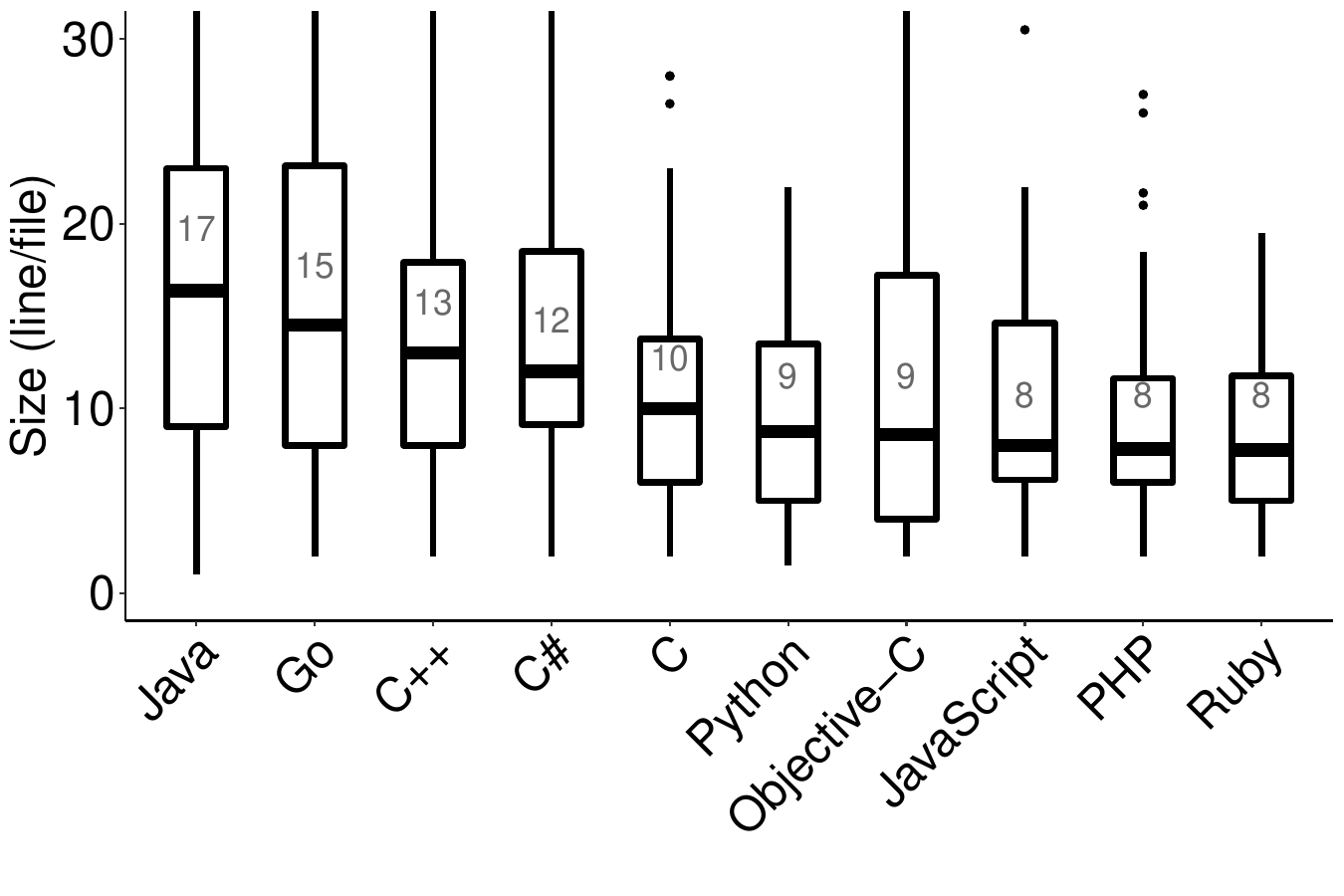}
 \end{tabular}}
        \vspace{-1mm}
    \caption{Extended Analysis: Aggregated/Normalised bug-resolution size for different languages. }
    \label{fig:combinemetric}
\end{figure}

\begin{figure}[t]
    \center
     \resizebox{.3\textwidth}{!}{
    \begin{tabular}{cc}
    \includegraphics[width = 0.44\linewidth,totalheight = 0.11\textheight]{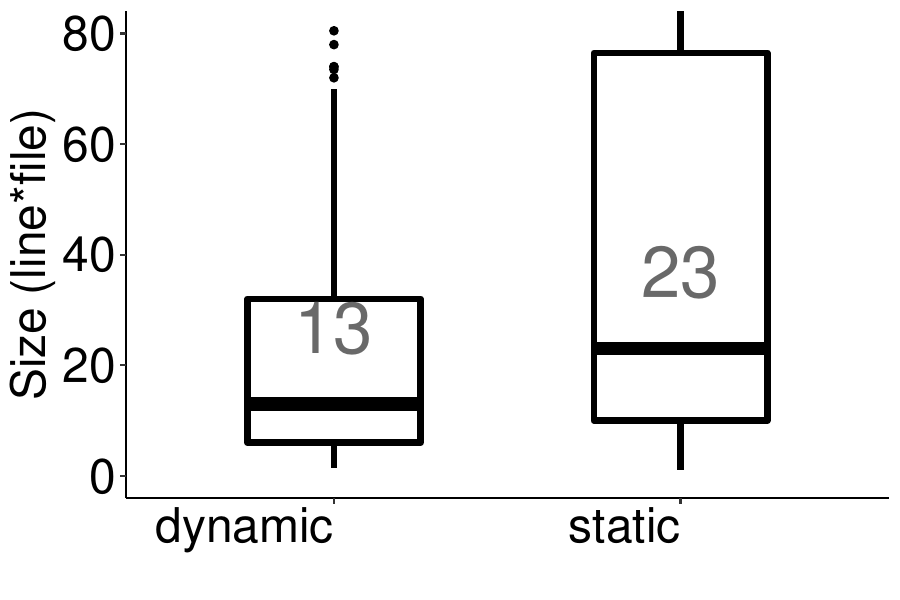}
    \includegraphics[width = 0.44\linewidth,totalheight = 0.11\textheight]{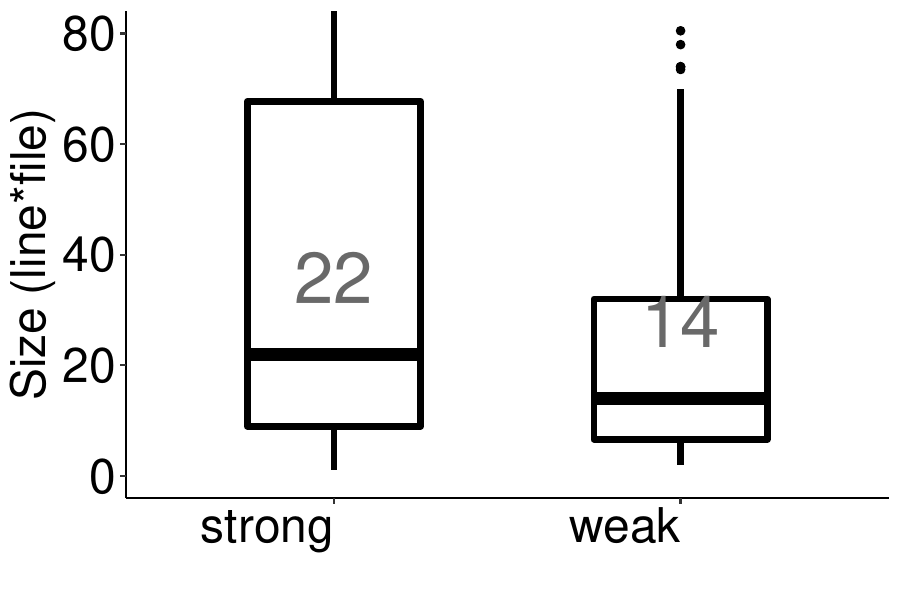}\\
    \includegraphics[width = 0.44\linewidth,totalheight = 0.11\textheight]{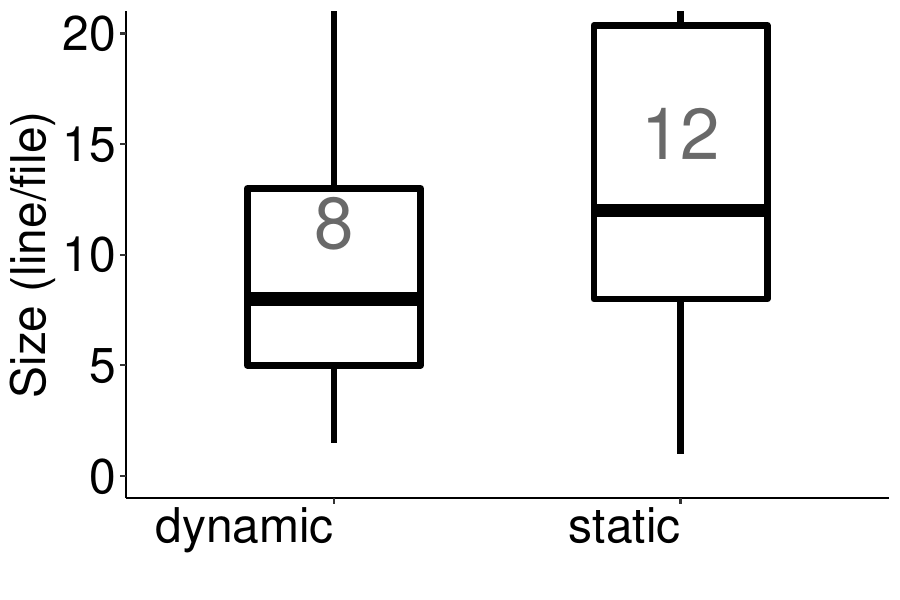}
    \includegraphics[width = 0.44\linewidth,totalheight = 0.11\textheight]{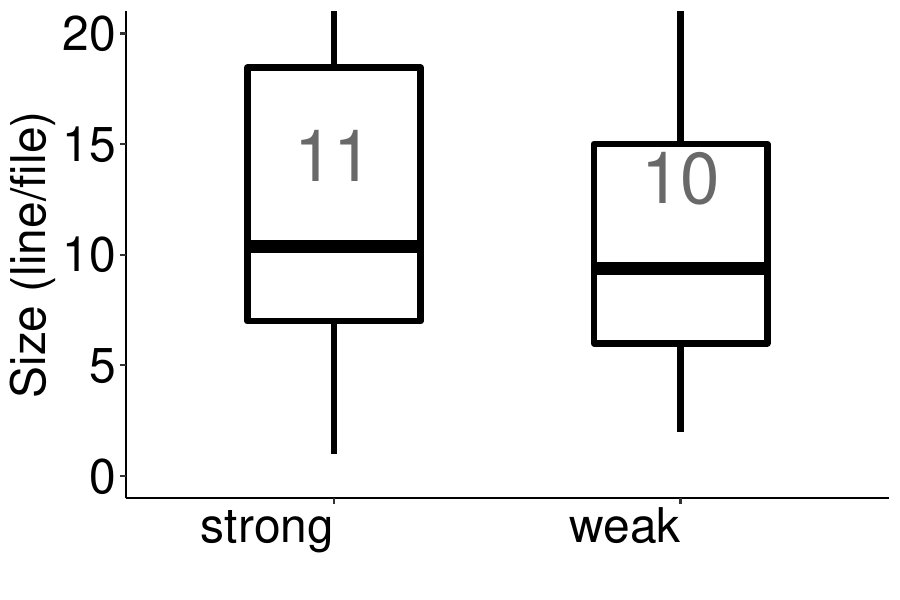}
    \end{tabular}}
    \caption{Extended Analysis: Aggregated bug-resolution size for different language categories.}
    \label{fig:combinemetriccategory}
\end{figure}

Figures~\ref{fig:combinemetric} and~\ref{fig:combinemetriccategory} show the results. 
Compared with Figures~\ref{fig:median} and~\ref{fig:mediancategory}, we see similar results for both languages and categories, except that the differences revealed by aggregated bug-resolution size are more obvious. 
This provides some evidence of robustness under different re-analyses, 
but more work is needed to provide greater certainty in our 
scientific conclusions from these observations.

\subsection{Results on Commits Not Related to Bug-Resolution}
\label{sec:non-buggy}
\pleasecheck{
This paper is about bug-resolution characteristics, and thus for data analysis we collected and focused on commits that are identified as bug-fixing. Nevertheless, the differences we have observed among the projects written in different languages/categories might have arisen from the language characteristics themselves. In other words, some language characteristics, e.g., the observed verbosity of \java{} projects, may also be observed for other commits that are not related to bug-resolution.}

\begin{figure}[h]
	\center
	 \resizebox{.3\textwidth}{!}{
	\begin{tabular}{c}
		\vspace{-3mm}
		\includegraphics[width = 0.76\linewidth,totalheight = 0.18\textheight]{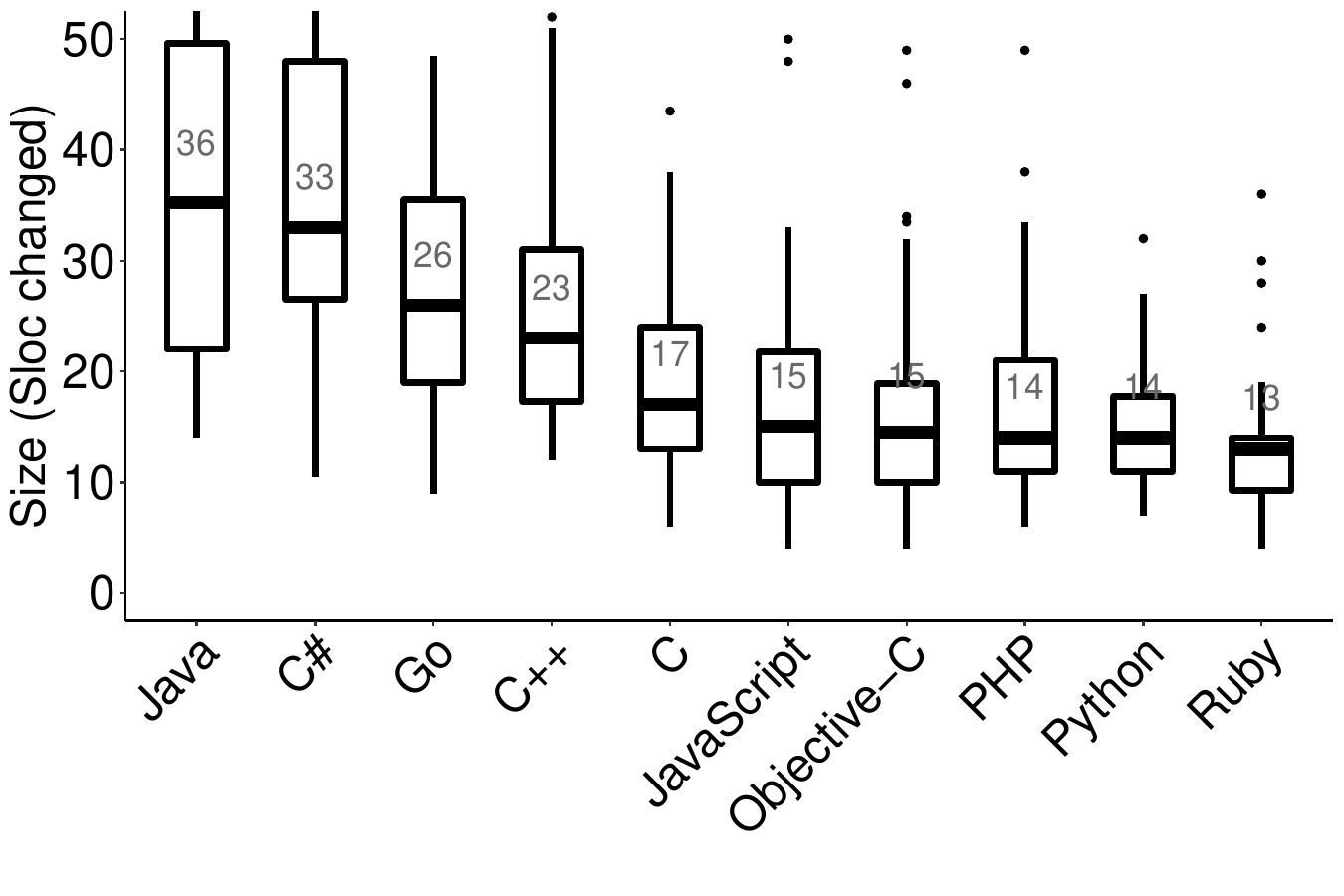}
		\\	
		\vspace{1.5mm}   \includegraphics[width = 0.76\linewidth,totalheight = 0.18\textheight]{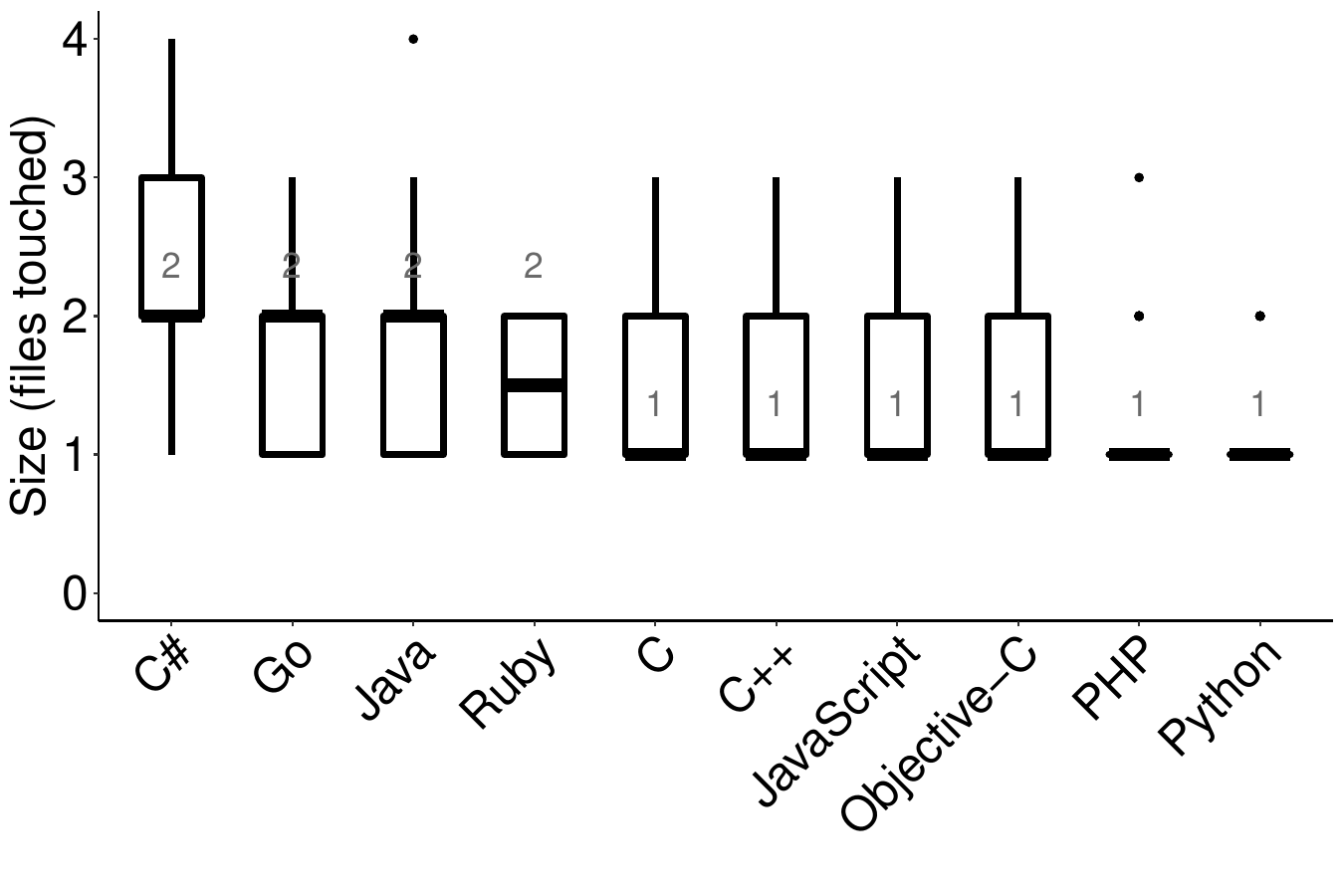}
		\\	
		\hspace{-3.5mm}   \includegraphics[width = 0.76\linewidth,totalheight = 0.18\textheight]{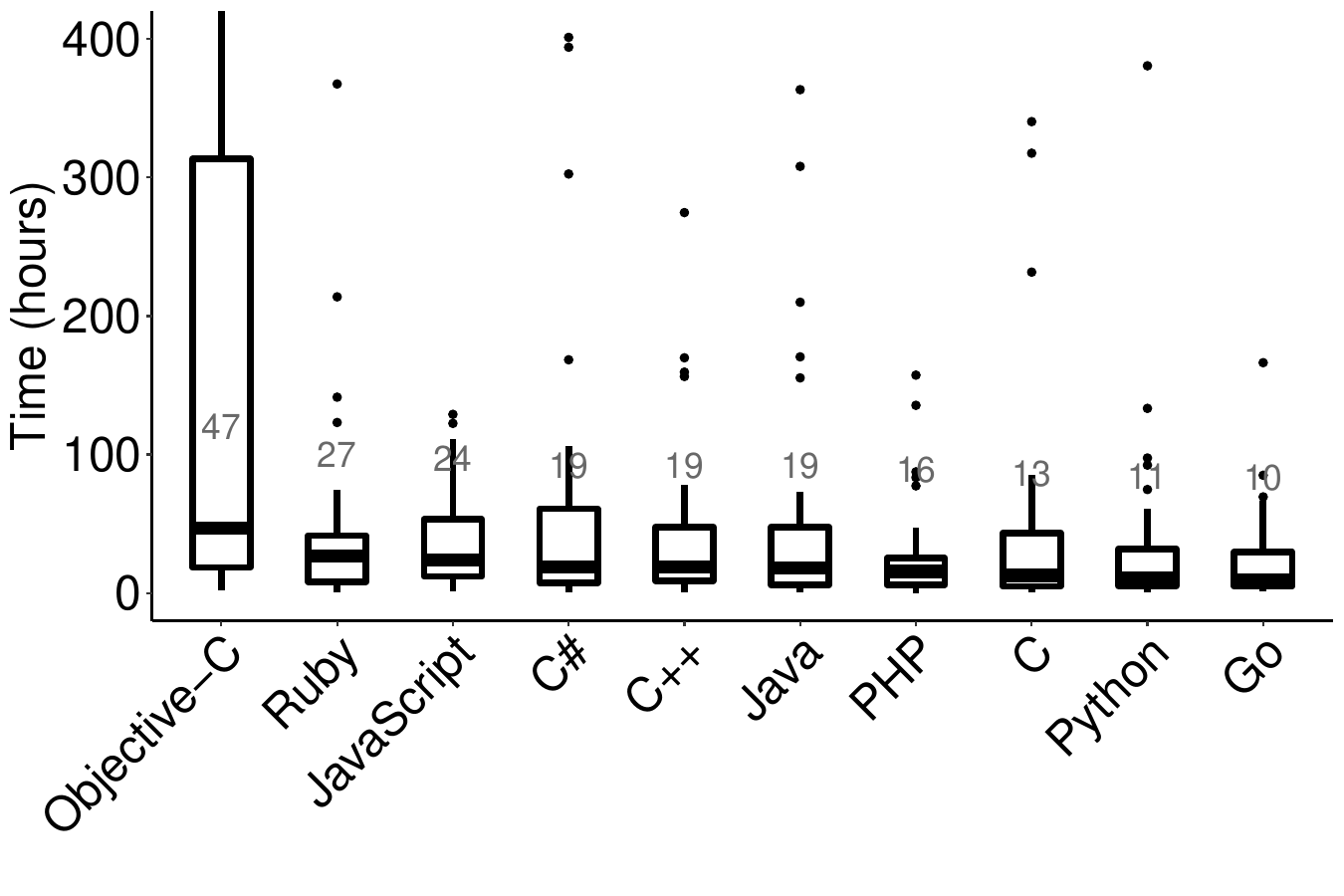}
		\\
	\end{tabular}}
	\vspace{-1mm}
	\caption{Extended analysis: Median-value comparison for non-buggy commits (project-level). }
	\label{fig:mediannonbuggy}
	\vspace{-3mm}
\end{figure}

To evaluate whether the observed differences exist in all commits, we repeat our project-level median-value analysis on the commits of each projects that are not identified as related to bug-resolution\footnote{We extract 1\% of the non-bug-resolution commits for each project due the large numbers.}. 
Figure~\ref{fig:mediannonbuggy} shows the results. 
Compared with Figure~\ref{fig:median}, we have the following observations for the projects in our corpus: 1) the ranking of languages based on the sizes of patches (i.e., SLOC and number of files) among non-bug-resolution commits is similar to that among bug-resolution commits; 2) however the same is not true for the issue resolution time ranking. For non-bug-resolution commits in different languages, the issue resolution time are similar to each other, in contrast to the larger disparity we observed for bug-resolution issues.

\pleasecheck{
These observations indicate that, for bug-resolution size, the
characteristics do carry over from the general case of all commits to the specific case of bug-fixing commits;
yet Simpsons's paradox~\cite{blyth1972simpson} alerts us to the fact that we cannot automatically assume such a `carry over' from general to specific without such analysis. 
}

\section{Implications}
\label{implication}
We \jienew{have presented evidence that} bug-resolution characteristics are different among projects written in different programming languages and language categories \jienew{in our corpus}.
In this section, we look at how the results, if replicated, may be useful in other areas of software engineering research.

\subsection{Improving Predictive Models for Planning}
\label{sec:prediction}
One potential application of our results is in the prediction of bug-resolution size/time, a problem that has been recognised as difficult, but has broad practical benefits in software development~\cite{weiss2007long,kaur2014software}. 
There are two categories of predictions. One is to estimate the resolution size/time of a specific bug in a project~\cite{weiss2007long,zhang2013predicting}. 
For multi-language projects, bugs belonging to different languages may have different bug-resolution size/time. 
The other is to predict the general level of bug-resolution cost of a project, rather than a specific bug~\cite{hayes2005maintainability,wohl1982maintainability}. 

As far as we are aware, no work in this area has considered programming paradigm. 
If these results are replicated, future work  might develop bug-resolution predictive models with the consideration of programming languages and paradigm, for better tuning and awareness of differences, should these prove more general, longstanding and widespread.

\subsection{Implications for Managers}
\label{sec:implicationsfordm}

Scheduling of tasks, including the task of bug resolution, is a major part of the software engineering process. 
Our results suggest that the bugs of different language ecosystems have different handling characteristics, which could be factored into the effort-estimation process. This is particularly true for multi-language projects, where one may need to consider the language attribute of each bug when assigning them.

\subsection{Implications for Researchers}
\label{sec:implicationforresearchers}

Our results suggest the possibility of including programming language as a feature in automatic bug-resolution prediction.
Our results indicate that such language-aware models may be more accurate for automatic prediction. For the area of automatic bug repair, 
our findings provide evidence that different languages may need patches of different sizes. 
Judged by the amount of line and file modification required, our results suggest that larger patches may be considered for automatically fixing for \csharp{}, \java{}, and \go{}. 
These languages may also benefit from techniques tuned to search a larger space (across more lines and files) for finding suitable patches.

Moreover, \jienew{program sets that exhibit higher than average bug-resolution time may require different approaches to those that exhibit lower resolution time.}
For the area of mutation testing, our results also suggest that projects in different languages may need different-size mutants to simulate real faults.

\section{Threats to Internal/External Validity}
\label{threats}

The primary threat to internal validity lies in the implementation of the study. To reduce this threat, the authors independently reviewed the experimental scripts to check their correctness.

The threats to external validity lie primarily with the subjects. We decided to pick the most popular projects for each language which, by definition, is not representative of all programs. However, we believe that it is more useful to study the most widely engaged and supported efforts of the communities, compared to, for example, randomly selecting projects.
Random selection may risk `polluting' the data with non-serious projects.

We took several steps to address \jienew{the threats to} construct validity.

\emph{Large dataset and multiple measurement metrics.} \jienew{Our experiment is relatively large scale, and we employed a variety of metrics to access the bug-resolution characteristics. }

\emph{Data validation.} 
To increase confidence and check for the threats to construct validity, we sampled a random selection of the data we collected, involving 520 commits from all selected projects, and manually checked them. We found that 85.4\% of them are `clean' (i.e., involving only the fixing of a single bug, and all the code modification is related to the bug-resolution).

To reduce the threat of language identification for a source code file, we use Github's own file extension library, the Linguist library, to identify relevant changed files. As far as we know, the file extensions in the Linguist library for the 10 languages we studied do not have coincidences that may bias language file identification.

Other threats to construct validity might emerge from the assessment of elapsed time in resolving a bug. We used the interval between the opening time and the time of the last comment before closing to approximate bug-resolution time. This has been shown to be a more accurate measurement of bug-resolution time than the (seemingly) more intuitive choice of the interval between the opening and closing time~\cite{Zheng:2015:MIC:2786805.2786866}.

\emph{Multiple analysis approaches.} To reduce the risk of bias caused by a single analysis approach, we refined and augmented the approach of Ray et al.~\cite{ray2014large} and adopted three different analysis approaches: multiple-regression analysis, median-value analysis, and the ScottKnott analysis. The consistency in the results of our different analyses tends to increase confidence. 

An inherent threat to validity in large-scale empirical studies such as ours derives from the potential for confounding factors. To reduce the number of potential independent variables that might otherwise confound our results, we investigated several non-language factors.

\section{Related Work}
\label{sec:relatedwork}
\textbf{Comparison of Languages on bug-resolution.}
Several previous authors have focused on the empirical comparison of either maintainability or bug-resolution characteristics among two or three types of programming languages. Bhattacharya et al.~\cite{bhattacharya2011assessing} statistically analysed four open-source projects developed in \cc{} and \cpp{}. They measured maintainability by the number of lines modified during bug-resolution, reporting that a move from \cc{} to \cpp{} results in reduced maintenance effort. The paper concluded that \cpp{} code requires less maintenance effort than \cc{}. 

Kleinschmager and Hanenberg et al.~\cite{kleinschmager2012static,hanenberg2014empirical} compared the bug-resolution time for \java{} and Groovy on one programming task. Their results indicate that programs written in Groovy (a dynamic languages) require more time in bug resolution, and attributed the difference to the benefit of static typing. 
Steinberg~\cite{steinberg2011impact} provided some evidence  that static typing exhibits some correlation with lower debugging time if only non-type errors are considered.   

Some researchers have written polemics against the usage of static languages. Nierstrasz et al.~\cite{nierstrasz2005revival} described static languages as ``the enemy of change'', claiming that dynamic languages are easier to maintain. Tratt et al.~\cite{tratt2009dynamically} also claimed that, compared to dynamic languages, static languages have higher development cost and require more complex changes. Sanner et al.~\cite{sanner1999python} described \python{} as a ``smaller, simpler, easy to maintain, and platform independent'' language due to its dynamic typing features. Oliphant et al.~\cite{oliphant2007python} gave a similar verdict.

\noindent\textbf{Comparison of Languages on Other Aspects.}
There is work comparing programming languages from other aspects, particularly software ``quality'' (i.e., the number of bugs generated rather than the effort required to handle them).

Phipps~\cite{phipps1999comparing} conducted an experiment to compare programmer productivity and defect rate for \java{} and \cpp{}, and reported that \java{} is ``superior''. %
Daly et al. \cite{daly2009work} empirically compared programmer behaviours under the standard \ruby{} interpreter, and DRuby which adds static type checking to \ruby{}. They found ``DRuby's warnings rarely provided information about potential errors".
Hanenberg et al.~\cite{hanenberg2010experiment} conducted an empirical study on a static type system for the development of a parser. They reported that ``the static type system has neither a positive nor a negative impact on an application's development time''.

\section{Conclusion}
\label{sec:conclusion}
We presented a large-scale study to investigate the connections between programs categorised by programming language and bug-resolution characteristics. 
We found evidence that for projects written in \java{},  bug resolution consumes less time than other languages, while for those written in \ruby{}, bug resolution consumes more time; we also found that statically typed projects have fixes that occupy more lines and touch more files than dynamically typed ones.
We found no evidence for correlation between bug-resolution time and size, nor any evidence for correlation with size, age, commit number, nor with target domain.

Inherent in this kind of large scale empirical study, concerning multiple projects and languages, there are  many confounding factors, data sanitisation issues, and other threats to the validity of any scientific conclusions and claims. 
While we have been careful to capture and highlight as many potential threats as we can, undoubtedly others remain. 
Nevertheless, we believe that such large-scale empirical studies are worthwhile. 
In particular, where observations reveal \pleasecheck{large median differences} in the corpora studied, this may suggest worthwhile avenues for re-analysis, replication and follow-on research.

\section*{Acknowledgement}
This work is supported by the ERC advanced grant with No. 741278, 
the National Key Research and Development Program of China under Grant No. 2017YFB1001803,
the NSFC Grant with No. 61861130363,
Newton Advanced Fellowships NAF$\backslash$R1$\backslash$180142, and
Royal Society IES$\backslash$R3$\backslash$170104 and NAF$\backslash$R1$\backslash$180142.